\documentclass[twocolumn,showpacs,aps,prx,longbibliography]{revtex4-1}
\usepackage{graphicx}  
\usepackage{amssymb} 
\usepackage{natbib}
\usepackage{amsmath}
\usepackage{enumerate}
\usepackage{textcomp}
\usepackage{times}
\usepackage{mathtools}
\usepackage{epsfig}
\usepackage[colorlinks,linkcolor=blue,citecolor=blue,urlcolor=blue]{hyperref}
\usepackage{color}
\usepackage{graphicx}
\usepackage{tabularx}
\usepackage{soul}
\usepackage[mathscr]{euscript}
\usepackage{amssymb}
\usepackage[normalem]{ulem}
\usepackage{mathrsfs}
\usepackage{amsmath}
\usepackage{color}
\usepackage{bbold}
\usepackage{comment}
\usepackage{bm}
\usepackage{braket}
\usepackage{dsfont}
\usepackage{textcomp}
\usepackage{lipsum}
\usepackage{amsmath,amssymb}
\newcommand{\pt}{\partial}
\newcommand{\mb}{\mathbf}

\newcommand{\md}{\mathcal{D}}

\newcommand{\egap}{\bar{\varepsilon}_{\mbox{\tiny{gap}}}}
\newcommand{\tegap}{\tilde{\varepsilon}_{\mbox{\tiny{gap}}}}
\newcommand{\gap}{\varepsilon_{\mbox{\tiny{gap}}}}
\newcommand{\eL}{\varepsilon_{\Lambda}}

\newcommand{\sx}{\sigma_x}
\newcommand{\sz}{\sigma_z}

\newcommand{\mc}{\mathcal}
\newcommand{\tr}{\bar{\mb{r}}}
\newcommand{\trr}{\mb{r}}
\newcommand{\tta}{\tau}

\newcommand{\Psc}{\Psi^{\scalebox{0.7}{sc}}}
\newcommand{\Pbs}{\Psi^{\scalebox{0.7}{bs}}}
\newcommand{\Msc}{\mathcal{M}_{ij}^{\scalebox{0.7}{sc}}}
\newcommand{\Mbs}{\mathcal{M}_{ij}^{\scalebox{0.7}{bs}}}
\newcommand{\Msci}{\mathcal{M}_{ii}^{\scalebox{0.7}{sc}}}

\newcommand{\Mscx}{\mathcal{M}_{xx}^{\scalebox{0.7}{sc}}}
\newcommand{\ems}{\varepsilon_{\mbox{\tiny{MS}}}}

\newcommand{\psc}{\psi^{\scalebox{0.7}{sc}}}
\newcommand{\Vsc}{\mathcal{V}_{m}^{\scalebox{0.7}{sc}}}
\newcommand{\Pfr}{\Psi^{\scalebox{0.7}{free}}}
\newcommand{\pfr}{\psi^{\scalebox{0.7}{free}}}
\newcommand{\Uwkb}{U_{\mbox{\tiny{WKB}}}}
\newcommand{\Mbsc}{\mathcal{M}_{ij}^{\scalebox{0.7}{bs-sc}}}
\newcommand{\Mbsci}{\mathcal{M}_{ii}^{\scalebox{0.7}{bs-sc}}}
\newcommand{\Mbscx}{\mathcal{M}_{xx}^{\scalebox{0.7}{bs-sc}}}
\newcommand{\Scl}{S^{\mbox{\tiny{WZ}}}_\textrm{cl}}
\definecolor{mygreen}{rgb}{0,0.65, 0}

\begin{document}
\title{Quantum Dynamics of Skyrmions in Chiral Magnets}
\author{Christina Psaroudaki$^{1}$, Silas Hoffman$^{1}$, Jelena Klinovaja$^{1}$, and Daniel Loss$^{1}$}
\affiliation{$^1$Department of Physics, University of Basel, Klingelbergstrasse 82, CH-4056 Basel, Switzerland}
\date{\today}
\begin{abstract}
We study the quantum propagation of a skyrmion in chiral magnetic insulators by generalizing the micromagnetic equations of motion to a finite-temperature path integral formalism, using field theoretic tools. Promoting the center of the skyrmion to a dynamic quantity, the fluctuations around the skyrmionic configuration give rise to a time-dependent damping of the skyrmion motion. From the frequency dependence of the damping kernel, we are able to identify the skyrmion mass, thus providing a microscopic description of the kinematic properties of skyrmions. When defects are present or a magnetic trap is applied, the skyrmion mass acquires a finite value proportional to the effective spin, even at vanishingly small temperature. We demonstrate that a skyrmion in a confined geometry provided by a magnetic trap behaves as a massive particle owing to its quasi-one-dimensional confinement. An additional quantum mass term  is predicted, independent of the effective spin, with an explicit temperature dependence which remains finite even at zero temperature. 
\end{abstract}

\maketitle
\section{Introduction}
\label{sec:intro}
\par
Although skyrmions have been proposed\cite{skyrmeNP62,Rossler06} long ago, there has been a strong rise in interest in recent years spurred  by experimental observations of skyrmionic phases in various  magnetic thin films\cite{Muhlbauer09,neubauerPRL09,pappasPRL09}. Magnetic skyrmions are characterized by a topologically nontrivial mapping from a two-dimensional (2D) magnetic system in real space into  three-dimensional spin space. 
This mapping leads to a  topological charge $Q$ which characterizes the skyrmion and is given by
\begin{equation}
Q=\frac{1}{4 \pi} \int d \mb{r}~ \mb{m} \cdot (\pt_x \mb{m} \times \pt_y \mb{m}) \,,
\end{equation}
where $\mb{m}$ is the normalized magnetization vector field and $x$ and $y$ are the spatial coordinates of the 2D magnetic layer\cite{Wilczek83,Papanicolaou91}. Magnetic skyrmions are attractive candidates for magnetic storage of classical information\cite{Fert13,kiselevJPD11} because they are topologically stable in the sense that no continuous local deformation in the magnetic texture can remove a skyrmion ({\it i.e.}, change $Q$), and they can be manipulated at high speed with relatively low current densities\cite{Jonietz10,iwasaki1NATcom13,Fert13}. Furthermore, because skyrmions coupled to conventional superconductors are known to support Majorana fermions\cite{yangPRB16},  they may be a route to topological quantum computation\cite{nayakRMP09}.

The dynamics of any classical magnetic texture with fixed  magnitude, $\mb{m}$, is governed by the Landau-Lifshitz-Gilbert (LLG) equation\cite{lifshitzBK80,gilbertTM04}. The dynamic properties of skyrmions can be considerably simplified by considering the motion of the average magnetic texture\cite{Thiele}, which reduces to an equation of motion for the skyrmion's center-of-mass coordinate, known as Thiele's equation. This skyrmion coordinate behaves as a massless particle under a Magnus force proportional to $Q$\cite{Stone96}, and a possible damping is parameterized by a phenomenological velocity-dependent term that is induced by the coupling of the skyrmion motion to other degrees of freedom in the system such as electrons, phonons, magnons, etc.; the microscopic details of this coupling are typically left unspecified. Using this phenomenological approach, several generalized equations have been proposed\cite{SchuttePRB90,Makhfudz,Everschor12,Iwasaki2013} for a rigid magnetic structure in the presence of external static as well as rapidly changing forces. The thermal diffusion of a driven skyrmion has been studied within the classical framework by a numerical solution of the stochastic LLG equation\cite{SchuttePRB90}, where random fields are included to model the effects of thermal fluctuations.

However, for small skyrmions at the nanometer scale, the size of skyrmions in state-of-the-art experiments\cite{Heinze2011,Wiesendanger16,Moreau16,Woo16}, quantum fluctuations are expected to become relevant\cite{Rolan15} and the classical LLG equation is expected to break down. 
In a fully quantum mechanical treatment starting from the spin and other degrees of freedom present in the system the quantum dynamics of the skyrmion should emerge
naturally, giving rise for instance to possible mass and dissipation terms linked to the underlying microscopics. 

The goal of this work is to provide such a quantum treatment of  skyrmions in insulating magnetic films. Thereby we restrict ourselves to systems where only spin degrees of freedom are present and other sources such as phonons and itinerant electrons are assumed to be negligibly small (due to low enough temperatures) or  entirely absent, respectively.
By employing quantum field theory methods and the  Faddeev-Popov techniques for collective coordinates~\cite{Gervais75a,Gervais75b,Sakita,Rajaman,Faddeev}, applied to magnetic systems~\cite{BraunPR96,Kyriakidis99},
we derive the quantum dynamics of the skyrmion in a systematic way, thereby going beyond the classical limit. 
 
 Within this formalism the skyrmion position, which parametrizes the classical static solution of the Euler-Lagrange equations, is promoted to a time-dependent dynamical variable and a finite perturbation theory in terms of fluctuations around the skyrmion configuration is performed. The interaction of the skyrmion with these magnon modes gives rise to  dissipation and to a nonlocal damping kernel whose form we obtain in a closed form and can evaluate analytically in some regimes of physical interest. In particular, we are able to microscopically calculate the  contribution to the damping in the Thiele equation from the incoherent magnon modes which in general depends on the magnetic dispersion relation and thus on the details of the free energy of the system. 
 
 For asymptotic (imaginary) times, we find that the damping kernel becomes local in time and reduces to a simple mass term which we can calculate explicitly in the presence of nonuniform magnetic traps and spatially dependent exchange interactions for arbitrary temperatures. Specifically, we find that local defects and magnetic traps which break translational symmetry induce a finite mass, even at vanishingly small temperature. As a striking consequence, we find that in quasi-one-dimensional wire geometries the skyrmion acquires a finite mass. Such geometries are provided for instance by anisotropic magnetic traps, similar to linear tracks used for magnetic memory and logic devices \cite{Fert13,Zhang15,Zhou14}. Surprisingly, this shows that skyrmions in quantum confined geometries exhibit a fundamentally different dynamical behavior compared to the one in unconfined geometries.

The paper is organized as follows. In Sec.~\ref{sec.quant} we derive a coherent-state path integral for a generic magnetic texture, while in Sec.~\ref{sec: Skyrmion Mass}, we utilize this functional integral approach to calculate the damping experienced by such a magnetic texture as a result of the coupling to magnon fluctuations, and we also demonstrate that this damping reduces to a simple mass term. In Sec.~\ref{sec:Chiral Magnets}, we apply our formalism to a chiral magnet and show in Sec.\ref{subsec:Mass Zero T}, that only a spatially nonuniform perturbation to the magnetic texture induces a skyrmion mass at zero temperature. To complete the description we examine in Sec.\ref{subsec:Mass Finite T} the skyrmion mass at finite temperatures. We conclude with a discussion and some general remarks in Secs.~\ref{sec:Discussion} and Sec.~\ref{sec:Conclusions}. Technical details are deferred to the Appendices.

\section{Collective Coordinate Quantization}
\label{sec.quant}

We consider a thin magnetic insulator with normalized magnetization 
$\mb{m} (\tilde{\mb{r}})=[\sin\Theta(\tilde{\mb{r}})\cos\Phi(\tilde{\mb{r}}),\sin\Theta(\tilde{\mb{r}})\sin\Phi(\tilde{\mb{r}},\cos\Theta(\tilde{\mb{r}})]$ in polar coordinates, where the $z$ axis points along the smallest dimension of the magnet (for the connection to spins in a 2D lattice model, see below).
For sufficiently thin magnets, {\it i.e.}, thinner than the exchange length, the magnetization is constant along the $z$ axis and $\tilde{\textbf{r}}=(\tilde{\rho} \cos \phi,\tilde{\rho} \sin \phi)$ is a vector in the $xy$ plane. The Lagrangian of such a magnetic system is given by
\begin{equation}
L= \frac{ S N_A}{\alpha^2} \int d \tilde{\mb{r}} ~\dot{\Phi} (1-\Pi)
 +N_A \int d \tilde{\mb{r}}~ \mathcal F(\Phi,\Pi) \,,
 \label{lagr}
\end{equation}
where $S$ is the magnitude of the spin, $N_A$ is the number of magnetic layers along the $z$ axis \cite{NoteN_A}, $\alpha$ is the lattice spacing, $\dot\Phi$ is the time derivative of $\Phi(\tilde{\mb{r}},t)$ and $\Pi(\tilde{\mb{r}},t)\equiv\cos\Theta$ is canonically conjugate to $\Phi$. The first term in Eq.~(\ref{lagr}), {\it i.e.}, the Wess-Zumino or Berry phase contribution (using the north pole parametrization\cite{BraunPR96}), describes the dynamics of the magnetization while the second term is the free energy which, in this and the following section, we keep general. For reasons of simplicity we assume that the energy density  functional $\mc{F}(\Phi,\Pi)$ is translationally  invariant whereas the rotational invariance is broken. This is supported by the fact that, for the majority of the known models which can stabilize magnetic skyrmions, U(1) rotational symmetry is broken by either the chiral Dzyaloshinskii-Moriya interaction or by long-range magnetic dipolar interactions. Upon minimizing $L$, one recovers the Euler-Lagrange or, equivalently, the Landau-Lifshitz (LL) equations, which govern the classical motion of the magnetization $\textbf m(\tilde{\mb{r}},t)$. 

To investigate the quantum effects of such a magnetic texture, we employ a functional integral formulation in which the partition function is given by $Z=\int \md\Pi \md \Phi ~e^{-S_E}$, where $S_E=\int_0^{\tilde{\beta}} d\tilde{\tau}\mathcal L$ is the Euclidean action. Here, $\mathcal L$ is the imaginary time Lagrangian, which is obtained upon replacing time $t$ with imaginary time $\tilde\tau=it$, and where $\tilde\beta=1/k_B\tilde T$ for the system temperature $\tilde T$ with  $k_B$ the Boltzmann constant. Throughout this work we assume that $\hbar=1$. We also introduce dimensionless variables $\mb{r}= \tilde{\mb{r}} / l\alpha$, $\tau= \tilde{\tau} \eL $, and $\beta=\tilde{\beta} \eL$.  Here $l$, a dimensionless constant, and $\eL$, a characteristic energy scale, depend on the system under study and are specified later in Sec.~\ref{sec:Chiral Magnets}.  
 With increasing number of layers  the effective spin $N_AS$ of the texture increases, and thereby $\mb{m}(\mb{r},\tau)$ approaches a macroscopic description of a classical magnetization field. We therefore consider the limit $N_A \gg 1$ to capture the leading semiclassical effects. 

\begin{figure}[!t]
\includegraphics[width=10cm]{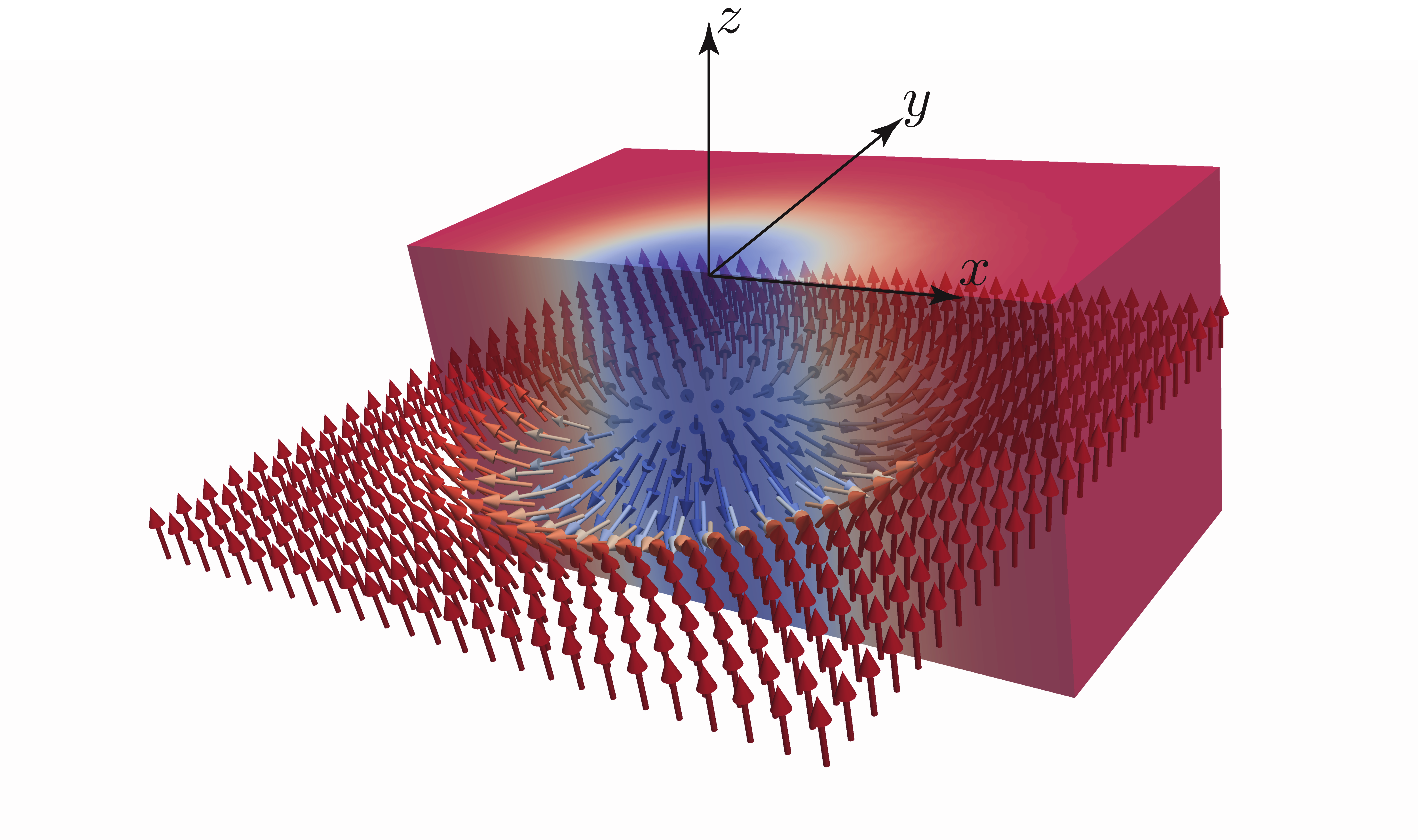}
\caption{\ Illustration of a  magnetic layer hosting a skyrmion with topological number $Q=1$ embedded in a bulk magnetic insulator. The magnetization  $\mb{m} (\tilde{\mb{r}})$ is assumed to be uniform along the  direction perpendicular to the layer. }
 \label{SkyrmionLayers}
\end{figure}

Henceforth, we consider a specific class of free energies which furnish a static stable or metastable solution, $\mb{m}_0 (\mb{r})=[\sin\Theta_0(\mb{r})\cos\Phi_0(\mb{r}),\sin\Theta_0(\mb{r}) \sin\Phi_0(\mb{r}),\cos\Theta_0(\mb{r})]$, in which $\mb{m}_0$ is characterized by a topological number $Q$ and the magnetization profile $\Theta_0$ approaches a constant value at spatial infinity $\vert \mb{r} \vert \rightarrow\infty$. These textures, known as Skyrmions, are illustrated in Fig.\ref{SkyrmionLayers} which depicts a two- dimensional magnetic layer hosting a Skyrmion with $Q=1$ embedded in a bulk magnetic insulator. We first develop a formalism that is valid for general Skyrmionic solutions, while an explicit solution of a Skyrmion is discussed in detail later. 

The center of mass of such solutions $\mb{m}_0 (\mb{r})$ is denoted by the collective coordinate $\mb{R}(\tau)$, which is energy independent owing to the translational symmetry of the system. The dynamics of the skyrmion is then described by $\mb{R}(\tau)$ and the magnetization fluctuations that arise when the
skyrmion moves. Because of the nonlinear character of the LL equation these fluctuations couple back to $\mb{R}(\tau)$ and thus affect the skyrmion motion.
To describe this effect we perform a canonical transformation of the path integral variables which separates the collective coordinate $\mb{R}(\tta)$ from the fluctuations around the skyrmion \cite{Gervais75a,Gervais75b},
 \begin{align}
\Pi(\mb{r},\tta)&=\Pi_0(\trr-\mb{R})+\eta(\trr-\mb{R},\tta)\,, \nonumber \\
\Phi(\mb{r},\tta)&=\Phi_0(\trr-\mb{R})+\xi(\trr-\mb{R},\tta)\,. 
\label{Transformation}
\end{align}

In the presence of translational symmetry there is a degenerate pair of zero-frequency modes $\xi  \propto \pt_{i} \Phi$ and $\eta  \propto \pt_{i} \Pi$, associated with translation of the skyrmion position in the $i=x,y$ direction. Instead of fluctuations $\xi$ and $\eta$, it is more convenient to work with the spinors $\chi$ and $\chi^{\dagger}$ defined by
\begin{equation}
\chi =\frac{1}{2} \binom{\xi \sin \Theta_0+i \eta/ \sin\Theta_0 }{ \xi \sin \Theta_0-i \eta/ \sin\Theta_0} \,.
\label{SpinorNotation}
\end{equation}
We denote as $\mc{Y}_x$ and $\mc{Y}_y$ the degenerate pair of translational zero modes expressed in the spinor notation of Eq.~\eqref{SpinorNotation}. We now construct a finite perturbation theory in terms of fluctuations around $\textbf m_0(\mb{r}-\mb{R})$,  where the system is described from a translated spatial frame $\tr= \mb{r} - \mb{R}(\tau)$. To second order in $\chi$, the action $S_E=S_\textrm{cl}+S_\textrm{fl}+S_I$ separates into three parts. The first part describes the translational motion of  $\textbf m_0$,
\begin{align}
\Scl=& -i \bar{S} N_A  \int_0^{\beta} d\tau \int d\mb{\tr}\, \dot{\mb{R}} \cdot(1- \Pi_0) \nabla \Phi_0  \,,
\end{align}
where $\bar{S}= S l^2$ and where we ignore an overall constant from the configuration energy 
$(l \alpha)^2\int d\mb{\tr} \mathcal F(\Phi_0,\Pi_0)$. Note that $\Scl$ contains a gauge-dependent boundary term proportional to $-i \bar{S} N_A \int d\tau \int d\mb{r}~\dot{\mb{R}} \cdot \nabla \Phi_0 =-i \bar{S} N_A \pi \int d\tau ~\dot{\mb{R}} \cdot \mb{C}$, where $\mb{C}=(1/\pi) \int d\mb{r}~ \nabla \Phi_0$ is the skyrmion chirality vector. This topological term does not affect the classical equation of motion, but could in general have observable effects, for example, in the tunneling probability \cite{Loss92,BraunPR96}. The most common class of solutions of magnetic skyrmions has vanishing topological term with $\mb{C}=0$, thus becoming manifestly a gauge invariant theory.

The second part consists of the fluctuations $\chi(\tr,\tau)$ around $\textbf m_0 (\tr)$, which we can consider as spin waves or magnons, given by
\begin{align}
S_{\textrm{fl}}=N_A\, \chi^{\dagger}\cdot \mathcal{G}\chi\, .
\end{align}
Here we introduce the compact scalar product notation for operators and functions,
\begin{align}
\chi^{\dagger}\cdot \mathcal{G}\chi\ &\equiv \int d\tau d\tau' d \tr  d \tr' \, \chi^{\dagger} (\mb{\tr},\tau)\mathcal{G}(\mb{\tr}, \tau; \mb{\tr'},\tau') \chi (\mb{\tr'},\tau')\,,
\end{align}
where the scalar product in spinor space is left implicit. Further we define
\begin{align}
\mathcal G&= \bar{S}  \sigma_z \partial_{\tau} + \mathcal{H}   \,,\nonumber\\
\mathcal H&=\frac{(l\alpha)^2}{\eL}\delta_{\chi^\dagger}\delta_\chi\mathcal F |_{\chi=\chi^\dagger=0}  ,
\end{align}
where $\delta_\chi$ is the functional derivative with respect to the field $\chi({\mb{\tr}}, \tau)$, and
$\mathcal G$ is the magnon Green's function. 
Finally, the coupling between the center of mass $\mb{R}(\tau)$  and the spin waves $\chi(\tr,\tau)$ is given by
\begin{align}
S_I&=N_A (\chi^{\dagger}\cdot\mathcal{K} \chi + \mathcal{J}^{\dagger}\cdot\chi+ \chi^{\dagger}\cdot \mathcal{J}), 
\end{align}
where
\begin{align}
\mathcal{K}&=   -\bar{S} \sz  \dot{\mathrm{R}}_i \Gamma_i \, , \nonumber\\
\mathcal{J}&=-\bar{S} \sz \dot{\mathrm{R}}_i f_i  \,.
\label{OperatorKJ}
\end{align}
Here, repeated indices, $i,j=x,y$, are summed over and  we introduce the abbreviations
\begin{align}
\Gamma_i&=\pt_i -  \sx \cot \Theta_0 \pt_i  \Theta_0 \, ,\nonumber\\
f_i &=\frac{1}{2}\binom{ \sin \Theta_0 \pt_i \Phi_0 - i \pt_i \Theta_0}{ \sin \Theta_0 \pt_i \Phi_0 + i \pt_i \Theta_0 }\,.
\end{align}
into the partition function, where $\delta(\cdot)$ is the Dirac delta function. We then obtain
\begin{align}
Z&=\int \md \mb{R}\, e^{- \Scl}~ \tilde{Z}[\mb{R}]\,,  \nonumber \\
\tilde{Z}&=\int  \md  {\chi}^{\dagger} \md {\chi}  ~\delta(G_x) \delta(G_y) \mbox{det} (J_{\mb{G}})  e^{-S_\textrm{fl}-S_I}\,,
\end{align}
where $J_\textbf{G}=d \textbf G(\tau)/d\mb{R}(\tau')$ is the Jacobian of the coordinate transformation and is treated as an additional term in the action, adaptable for perturbative calculation. We should stress that the scalar product $\chi ^{\dagger} (\tr) \mc{Y}_i(\tr)$ (in the 2-spinor space) is determined by the properties of the Hamiltonian $\mc{H}$. In most cases $\mc{H}$ is not Hermitian for the standard metric and orthogonality conditions are enforced by an inner product of the form $\chi ^{\dagger} (\tr) \sz \mc{Y}_i(\tr)$  \cite{Sheka04}. 

We next integrate out the fluctuations $\chi$ and $\chi^{\dagger}$ from the partition function neglecting terms $O(1)$ in $N_A$ which originate from the FP determinant $\mbox{det} (J_\textbf{G})$. The integral over  $\chi$ and $\chi^{\dagger}$ can be reduced to a Gaussian form by completing the square, 
$\tilde{\chi} =\chi + (\mc{G}+\mc{K})^{-1} \mc{J} $.
The inverse of an operator, $O^{-1}$, is defined by $\int d\tr'' d\tau'' O(\tr,\tau; \tr'',\tau'')O^{-1}(\tr'',\tau''; \tr',\tau')=\delta(\tr-\tr')\delta(\tau-\tau')$.
After integration, the partition function reduces to
\begin{align}
\tilde{Z}[\mb{R}] &= \int  \md  \tilde{\chi}^{\dagger} \md \tilde{\chi} \delta(G_x) \delta(G_y) e^{-\mc{I} - N_A  \tilde{\chi}^{\dagger}\cdot (\mc{G}+\mc{K})\tilde{ \chi}  } \nonumber \\
&= e^{-\mc{I}} \frac{1}{\mbox{det}'[N_A(\mc{G}+\mc{K})] }\,,
\label{ActionDeterm}
\end{align}
where the prime on the determinant excludes the zero modes, and where
\begin{align}
\mc{I}&=-N_A \mc{J}^{\dagger}\cdot (\mc{G}+\mc{K})^{-1} \mc{J} \nonumber\\
&\approx-N_A \mc{J}^{\dagger}\cdot \mc{G}^{-1} \mc{J} \,.
\label{apI}
\end{align}
Upon exponentiating the determinant and expanding to lowest nonvanishing order in $\dot{\mb{R}}$, we get
\begin{equation}
\frac{1}{\mbox{det}'[N_A(\mc{G}+\mc{K})] }\approx \frac{e^{- \textrm{Tr}'[- \frac{1}{2}(\mc{G}^{-1} \mc{K})^2]} }{\mbox{det}'(N_A \mc{G})}\,,
\label{apDet}
\end{equation}
where $\textrm{Tr}' O=\int d\tau d{\tr} O(\tr,\tau;\tr,\tau)$ denotes the trace and where
we use that $\textrm{Tr}'(\mc{G}^{-1}\mc{K})=0$ due to periodic boundary conditions in time. The approximations in Eqs.~(\ref{apI}) and Eq.~(\ref{apDet}) neglect terms $\mathcal O( \dot{\mb{R}}^3)$. 
 
The quantization of a skyrmionic field presented in this section, with the dynamical part of the action proportional to $\dot{\Phi} \Pi$, is significantly different from the canonical quantization of theories with a dynamical part proportional to $\dot{\Phi}^2$, for example the quantization of a kink soliton of a (1+1) nonlinear field theory\cite{Gervais75a} or a real two-component scalar field in (2+1) dimensions \cite{Dorey94}.

\section{Skyrmion Mass and Dissipation}
\label{sec: Skyrmion Mass}
Because $\tilde Z[\mb{R}]$ is, in general, nonlocal in time, it naturally encodes dissipation of the skyrmion that is captured by the action
\begin{align}
S_d=-N_A \mc{J}^{\dagger}\cdot \mc{G}^{-1} \mc{J}-\textrm{Tr}'[\frac{1}{2}(\mc{G}^{-1} \mc{K})^2]\,.
\label{DissAction}
\end{align} 
Let $\Psi^{s}_n(\tr) e^{i\omega_\nu\tau}/\sqrt{\beta}$ be the normalized eigenvectors of the operator $\mc G$, 
\begin{equation}
\mc G\, \Psi^{s}_n(\tr) e^{i\omega_\nu\tau}= (\bar{S} i\omega_\nu  +2  \varepsilon^{s}_n) \sz  \Psi^{s}_n(\tr) e^{i\omega_\nu\tau}, 
\label{eigenvector}
\end{equation}
where  $\omega_\nu=2 \pi \nu / \beta$  are the Matsubara frequencies, with $\nu$ integer, and where $\varepsilon^{s}_n$ is the corresponding eigenenergy of the magnons, given as solutions of the eigenvalue problem (EVP) $\mc{H} \Psi^{s}_n= 2 \varepsilon^{s}_n \sz \Psi^{s}_n$. Here we introduce index $s=\pm 1$ to distinguish between particle states $\Psi^{1}_n(\mb{r})$ with eigenfrequency $\varepsilon^{1}_{n} = + \varepsilon_{n}$ and antiparticle states $\Psi^{-1}_n(\mb{r})$ with eigenfrequency $\varepsilon^{-1}_{n} = - \varepsilon_{n}$. We refer the reader to Appendix \ref{sec:ap2} for a detailed discussion. Using this complete basis we get
\begin{equation}
\mc{G}^{-1}(\tr,\tau;\tr',\tau')=\frac{1}{\beta} \sum_{\substack{i\omega_{\nu},n \\ s=\pm 1}}^{}{'}  \frac{s \sz \Psi^{s}_n(\tr) (\Psi^{s}_n(\tr'))^\dagger \sigma_z e^{i\omega_\nu(\tau-\tau')}}{\bar{S} i\omega_\nu  +2  \varepsilon^{s}_n} \,.
\label{InverseG}
\end{equation}
 Insertion of the series expression \eqref{InverseG} into Eq.~\eqref{DissAction} yields
\begin{align}
 S_d=\int_0^{\beta} d\tau  \int_0^{\beta} d\sigma~ \dot{\mathrm{R}}_i(\tau) \gamma_{ij}(\tau-\sigma) \dot{\mathrm{R}}_j(\sigma)  \,,
\label{DissKernel2}
\end{align}
where summation over repeated indices $i,j=x,y$ is implied. The damping kernel $\gamma_{ij}=\gamma_{ij}^{0}+\gamma_{ij}^{T}$  consists of a term proportional to the effective spin $N_A S$,
\begin{align}
\gamma_{ij}^{0}(\tau) =  N_A\bar{S}\frac{1}{\beta} \sum_{s=\pm1} \sum_{i \omega_\nu} \sum_{n}^{}{'} \frac{ s \mc{A}_{ij}^{n,s} ~e^{i \omega_\nu \tau}}{i\omega_{\nu} + \bar{\varepsilon}^{s}_n} \,,
\label{DissClas}
\end{align}
where $\bar{\varepsilon}^{s}_n= 2 \varepsilon^{s}_n/\bar{S}$ with $\bar{S}= S l^2$, the prime in the sum denotes omission of the zero modes, and the elements $\mc{A}_{ij}^{n,s}$ are given by
\begin{align}
\mc{A}_{ij}^{n,s}=  \langle f_i, \Psi^{s}_n \rangle \langle  \Psi^{s}_n, f_j \rangle \,.
\label{ElementA}
\end{align}
The second part of the kernel is independent of the effective spin $N_A\bar{S}$ and is calculated by evaluating the trace in Eq.~\eqref{DissAction} in the eigenbasis of the operator $\mc{G}$,
\begin{align}
\gamma_{ij}^{T}(\tau) =  \frac{1}{\beta^2} \sum_{s,s'=\pm 1} \sum_{i \omega_\nu,i \omega_{\nu'}} \sum_{n,n'}^{}{'} \frac{ ss'\mc{B}_{ij}^{ns,n's'} e^{i (\omega_\nu -\omega_{\nu'})\tau}}{(i\omega_{\nu} + \bar{\varepsilon}^{s}_n)(i\omega_{\nu'} + \bar{\varepsilon}^{s'}_{n'})} \,,
\label{DissQua}
\end{align}
with matrix elements
\begin{align}
\mc{B}_{ij}^{ns,n's'}=\frac{1}{2} \langle \Psi^{s}_{n}, \sz \Gamma_i \Psi^{s'}_{n'} \rangle \langle  \Psi^{s'}_{n'},  \Gamma_j \sz\Psi^{s}_{n} \rangle \,.
\label{ElementB}
\end{align}
The sums \eqref{DissClas} and \eqref{DissQua} over Matsubara frequencies $\omega_\nu$ can be explicitly performed using the following exact relation,
\begin{equation}
\frac{1}{\beta} \sum_{\nu} \frac{e^{i \omega_\nu \tau}}{\omega_\nu^2 + \varepsilon_n^2} = \frac{1}{2\varepsilon_n} \frac{\cosh (\varepsilon_n (\vert \tau \vert -\beta/2))}{\sinh(\beta \varepsilon_n/2)} \,,
\label{MatsuSum1}
\end{equation}
where  the RHS is understood to be periodically extended beyond $|\tau| \leq \beta/2$. The action Eq.~\eqref{DissKernel2} takes the typical form of a system coupled to a thermal reservoir governed by a linear dissipative process \cite{WeissBook}. The expression for $\mc{S}_d$ contains contributions that are nonlocal in imaginary time and describe effective correlations of $\mathrm{R}_i$ with itself at different times.
These correlations are a consequence of an emission of a virtual magnon at time $\tau$ generated by the motion of the skyrmion which is then reabsorbed  by the skyrmion
at a later time $\tau'$.  The contributions at equal times, if finite, will give rise to a mass of the skyrmion, which otherwise is absent. The form of the kernel $\gamma_{ij}$ obviously depends on the microscopic details of the system and determines the amplitude of these effects as see below.
\begin{table*}
\caption{\label{Units} Relation between physical (tilde) and dimensionless parameters}
\begin{ruledtabular}
\begin{tabular}{ccccccc}
 Length&Imaginary Time&Temperature&Magnetic Field&Easy-axis Anisotropy&Magnon Gap
\\ \hline
\\
 $\tilde{\mb{r}}= \mb{r}\frac{\tilde{J} \alpha}{\tilde{D}}$&$\tilde{\tau}= \tau \frac{\hbar}{\tilde{J} S^2} $&$\tilde{T}= T \frac{\tilde{J} S^2}{k_B}$&$g \mu_B \tilde{H} =h \frac{ \tilde{D}^2}{\tilde{J}}$&$\tilde{K}=\kappa \frac{ \tilde{D}^2}{\tilde{J}}$&$\tegap=\tilde{J} S^2 \egap$  \\
\end{tabular}
\end{ruledtabular}
\end{table*}

For the case of a translationally invariant free energy, the translational modes with zero energy are proportional to the functions $f_i$. Because  the summation over the magnon modes excludes translation, $\mathcal A_{ij}^{n,s} =0$, we conclude that there is no  mass contribution from the damping kernel $\gamma_{ij}^0$. This is in agreement with the well-known result that the skyrmion behaves like a massless particle in the presence of a magnetic field \cite{Thiele}. However, the situation changes if we allow for a perturbation in our free energy, denoted by $\mathcal V(\tr),$ which explicitly breaks translational symmetry. Treating $\mathcal V(\tr)$ in standard perturbation theory, the magnon states become in lowest order,
\begin{align}
\tilde \Psi^{s}_n(\tr)=\Psi^{s}_n(\tr)+\sum_{s'=\pm1} \sum_{m \neq \Psi^{s}_n}\frac{\Psi^{s'}_m(\tr)}{\varepsilon^s_n-\varepsilon^{s'}_m} \langle \Psi^{s'}_m,  \mathcal W \Psi^{s}_n \rangle \,,
\end{align}
where $\mc{W} =\delta_{\chi^\dagger}\delta_\chi\mathcal V |_{\chi=\chi^\dagger=0} $. Using the orthogonality properties of the unperturbed magnon modes, we find to lowest order in $\mathcal V(\tr)$,
\begin{align}
\mc{A}_{ij}^{n,s}=\frac{1}{(\varepsilon_n^s)^2}\langle f_i, \sz\mathcal W \Psi^{s}_n \rangle \langle \Psi^{s}_n, \mathcal W\sz f_j \rangle\,,
\end{align}
which in general is nonzero (see below). In contrast to $\gamma^0_{ij}$, the second contribution to dissipation, $\gamma^{T}_{ij}$, assumes a finite value in general, {\it i.e.}, we find nonzero matrix elements $\mc{B}_{ij}^{ns,n's'} \neq 0$, even for a translationally invariant system. 

The first variation of the action $S_{cl} + S_d$ vanishes for the extremal (classical) path $\mathrm{R}_i(\tau)$ obeying the following equation of motion in bosonic Matsubara $\omega_\nu$ frequency space,
\begin{align}
\tilde{Q} \omega_{\nu} \epsilon_{ij} \mathrm{R}_{j}^{\omega_{\nu}}+ \omega_{\nu}^2 (\gamma_{ij}^{\omega_{\nu}} + \gamma_{ji}^{-\omega_{\nu}})  \mathrm{R}_{j}^{\omega_{\nu}} = - V_i^{\omega_{\nu}}\,,
\label{sEoM}
\end{align}
where $\epsilon_{ij}$ is the Levi-Civita tensor and summation over repeated indices $i,j$ is implied.  Here, the $\beta$-periodic functions in imaginary time are expanded into Fourier series as
\begin{align}
\mathrm{R}_i(\tau)= \frac{1}{\beta} \sum_{i \omega_{\nu}} e^{i \omega_{\nu} \tau} \mathrm{R}_i^{\omega_{\nu}} \,,
\end{align}
and we assume the presence of a potential term $\mc{V}$ which breaks translational symmetry which is incorporated in the equation of motion
 through 
 \begin{align}
V_i^{\omega_{\nu}}=   \int_0^{\beta} d\tau~ e^{-i \omega_{\nu} \tau} \frac{\delta \mathcal V}{\delta \mathrm{R}_i} \,.
\end{align}
The first term in \eqref{sEoM} originates from the Berry phase and results in a Magnus force acting on the skyrmion proportional to the winding number $\tilde{Q}=4 \pi \bar{S} N_A Q$, recovering well-known results~\cite{Thiele,Stone96}. The Fourier coefficients $\gamma_{ij}^{\omega_{\nu}}$ are found from Eqs.~\eqref{DissClas} and \eqref{DissQua},
\begin{align}
&\gamma_{ij}^{\omega_{\nu}}=N_A \bar{S}\sum_{s=\pm1} \sum_n{}{'} \frac{s \mc{A}_{ij}^{n,s}}{i\omega_\nu +\bar{\varepsilon}^{s}_n} 
 \nonumber \\
&+\sum_{s,s'=\pm1}\sum_{n,n'}{}{'}\frac{s s'\mc{B}_{ij}^{ns,n's'}}{2} \frac{\coth( \frac{\beta \bar{\varepsilon}^s_{n}}{2}) - \coth( \frac{\beta \bar{\varepsilon}^{s'}_{n'}}{2} )}{\bar{\varepsilon}^{s'}_{n'} - \bar{\varepsilon}^s_{n} +i \omega_\nu} \,.
\label{DampFreq}
\end{align}
The quantum dynamics of the skyrmion is described by a kernel $\gamma_{ij}^{\omega_{\nu}}$ with a strong frequency dependence, in contrast to the typical Gilbert damping. However, to be consistent with the approximation that terms $\mathcal O( \dot{\mb{R}}^3)$ are negligible, we note that the damping kernel $\gamma_{ij}^{\omega_{\nu}}$ includes processes occurring at frequencies $\omega_\nu \lesssim \gamma_{ij}^{\omega_{\nu}}$ and the behavior for higher frequencies is beyond the range of validity of our derivation. 
For low frequencies, \textit{i.e.} asymptotic imaginary time $\tau$, the damping kernel $\gamma_{ij}^{\omega_\nu}$ in Eq.~\eqref{sEoM} can be expanded,
$\gamma_{ij}^{\omega_\nu} =\gamma_{ij}^{\omega_\nu=0}+ \mathcal{O}(\frac{\omega_{\nu}}{\egap}) $,
which holds for all characteristic frequencies in the range $\omega_{\nu} < \egap$, and consequently for temperatures in the range $T< \egap$, where $\egap$  is the lowest magnon energy gap. We therefore observe that the effect of the damping can be reduced to a mass term defined as 
\begin{align}
\mc{M}_{ij} =\mbox{Re} [\gamma_{ij}^{\omega_\nu=0}+\gamma_{ji}^{-\omega_{\nu}=0} ] \equiv \mc{M}_{ij}^{0} + \mc{M}_{ij}^{T} \,,
\label{MassTerm} 
\end{align}
where the first term, which is temperature-independent and proportional to the effective spin $N_A \bar{S}$, is given by
 \begin{align}
\mc{M}_{ij}^{0}=N_A \bar{S} \sum_{s=\pm1}\sum_n{}{'} \frac{s \mbox{Re}[\mc{A}_{ij}^{n,s}+\mc{A}_{ji}^{n,s}]}{\bar{\varepsilon}^s_n}  \,,
\label{Mass0}
\end{align}
while the second mass term is explicitly temperature dependent and independent of $N_A \bar{S}$,
\begin{align}
\mc{M}_{ij}^{T}=  \sum_{s',s=\pm1}\sum_{\substack{n,n' }} ss' \mbox{Re}[\mc{B}_{ij}^{ns,n's'}]\frac{\coth( \frac{\beta \bar{\varepsilon}^{s'}_{n'}}{2}) - \coth( \frac{\beta \bar{\varepsilon}^{s}_{n}}{2})}{ \bar{\varepsilon}^{s}_{n}- \bar{\varepsilon}^{s'}_{n'}} \,. \label{MassT}       
\end{align}
Note that there is a nonvanishing particle-antiparticle contribution to the effective mass at zero temperature,
\begin{align}
\mc{M}_{ij}^{T \rightarrow 0}=\sum_{n,n'}{}{'}\frac{\mbox{Re}[\mc{B}_{ij}^{n1,n'-1}+\mc{B}_{ij}^{n-1,n'1}]}{\bar{\varepsilon}_{n'}+ \bar{\varepsilon}_{n} } \,,
\end{align}
which reflects the presence of quantum fluctuations at zero temperature.

The definition of the mass as the zero-frequency limit of the real part of the damping kernel $(\gamma_{ij}^{\omega_{\nu}}+ \gamma_{ji}^{-\omega_{\nu}})$ is valid under the assumption that the real-time dynamics of the skyrmion is described by circular modes $\tilde{\omega}$ that are small compared to the magnon gap $\tilde{\omega} < \egap$. For example, in the presence of a parabolic potential well $\mc{V}= \mathrm{K} (\mathrm{R}_x^2+\mathrm{R}_y^2)$, the characteristic frequency of the circular motion in the absence of a mass is $\tilde{\omega}=\mathrm{K}/ \tilde{Q}$. Here, $\mc{V}$ is a perturbation to the magnon Hamiltonian $\mc{H}$ and $\tilde{Q}$ is proportional to the effective spin $N_A S$, thus the $i \omega_\nu \rightarrow 0$ limit in the mass definition is well justified. In the opposite case of a large dominant frequency, the mass tensor is refined around $\tilde{\omega}$ as $\mc{M}_{ij} =\mbox{Re}[ \gamma_{ij}^{ \omega_{\nu} = -i \tilde{\omega}}+\gamma_{ji}^{ \omega_{\nu} = i \tilde{\omega}}]$. 
Finally, note that ordinary friction, of the form of a damping proportional to the velocity of the skyrmion, is also included in our formulas and can be identified by the frequency-dependent coefficient $d_{ij}^{\omega_\nu}=- \omega_\nu~\mbox{Im}[\gamma_{ij}^{\omega_\nu}+\gamma_{ji}^{-\omega_\nu}]$. 
 
The equation of motion Eq.~(\ref{sEoM}) and the mass tensor Eq.~\eqref{MassTerm} are the main results of this section. They are valid for a general magnetic texture 
$\mb{m}(\mb{r},\tau)$ satisfying the specified requirements, and govern the dynamics of its center-of-mass position ${\bf R}(\tau)$, taking into account quantum and thermal fluctuations  and allowing for a potential $\mathcal V$ that weakly breaks the translational symmetry. This equation includes only dissipation produced by the magnon modes, while other sources giving rise to damping such as phonons and itinerant electrons are not included. Thus, we also must assume that we work at low enough temperatures such that phonon effects can be safely neglected. It is worth mentioning that additional forces should be included in Eq.~\eqref{sEoM} in the presence of a constant magnon current generated by a source \cite{SchuttePRB14} or by a magnon current induced by a temperature gradient \cite{Schroeter16}. 

The presence of a non-negligible mass term leads to additional oscillatory modes in the real-time dynamics of the skyrmion \cite{Makhfudz,Buttner15}. From the structure of the elements $\mc{A}_{ij}^{n}$ of Eq.~\eqref{ElementA} and $\mc{B}_{ij}^{n,n'}$  of Eq.~\eqref{ElementB}, it is easy to see that in the presence of a nonuniform but isotropic potential $\mc{V}$ or at finite $T$, the diagonal mass terms are equal, $\mc{M}_{xx}=\mc{M}_{yy}$, while off-diagonal terms vanish, $\mc{M}_{xy}= \mc{M}_{yx} = 0$, because the real part of matrix elements $\mc{A}^{n,s}_{ij}$ and $\mc{B}_{ij}^{ns,n's'}$ is antisymmetric in indices $i,j = x,y$. Therefore, in the absence of a driving force and assuming that the confinement is given by a parabolic well $\mc{V}= K (\mathrm{R}_x^2+\mathrm{R}_y^2)/2$, the real-time dynamics of the skyrmion is oscillatory with two characteristic circular modes given by
\begin{equation}
\tilde{\omega}_{\pm} =- \frac{\tilde{Q}}{2 \mc{M}} \pm \sqrt{ \left(\frac{\tilde{Q}}{ 2 \mc{M}}\right)^2 + \frac{K}{ \mc{M}} } \,,
\end{equation}
where we introduce $\mc{M}= \mc{M}_{xx}= \mc{M}_{yy}$. Finally, the most general case of an anisotropic potential $\mc{V}$ gives rise to four distinct circular modes, originating from the fact that the mass tensor elements are no longer equal, a result that could possibly have interesting experimental implications.
\begin{figure}[t]
\centering
\includegraphics[width=1\linewidth]{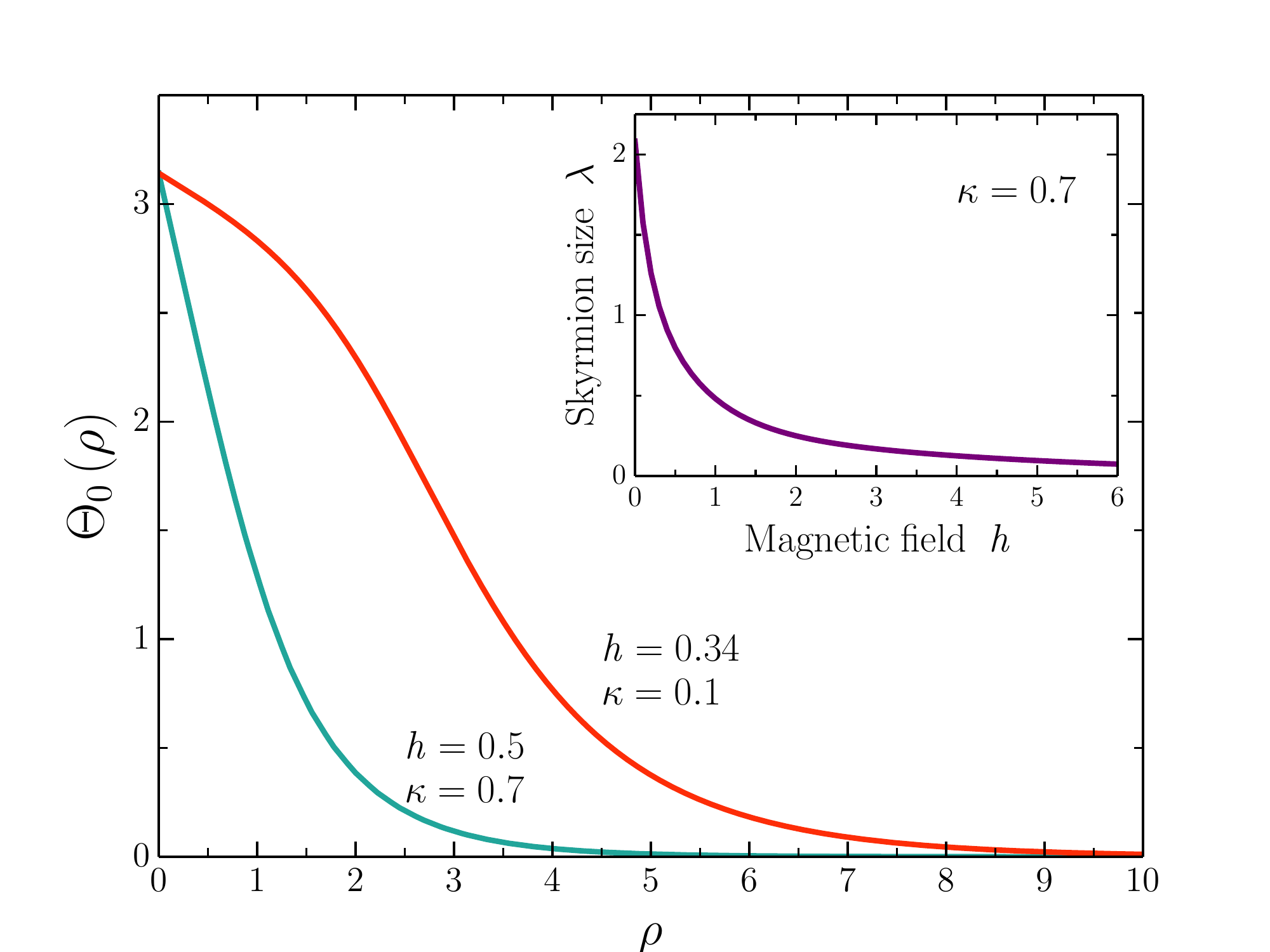}
\caption{\ Magnetization profiles $\Theta_0(\rho)$ of a skyrmion as function of radial distance $\rho$ for a small-radius skyrmion with chosen easy-axis anisotropy $\kappa=0.7$ and magnetic field $h=0.5$ (green line) and a large-radius skyrmion with $\kappa=0.1$ and $h=0.34$ (red line). Magnetic field is applied along the easy axis (all parameters given in scaled dimensionless units, see text).  The inset depicts the dependence of the skyrmion radius $\lambda$ on  $h$ for $\kappa=0.7$. 
} \label{Profile}
\end{figure}

\section{Chiral Magnets}
\label{sec:Chiral Magnets}
In this section, we apply the  formalism derived in the previous section explicitly to a skyrmionic magnetic texture, focusing on the low-temperature contribution to the mass, in the presence of a perturbation in the free energy that breaks  translational symmetry. Skyrmions, known to be stable or metastable solutions in chiral magnets with an applied magnetic field, are described on a square lattice by the spin Hamiltonian,
\begin{align}
H_s=  &-\tilde{J} \sum_{a,i}  \mb{S}_a \cdot \mb{S}_{a+e_i}  - \tilde{K} \sum_a (S^{z}_a)^2 - g \mu_B \tilde{H} \sum_a S^z_a  \nonumber \\
&-\tilde{D} \sum_{a}  \left[ (\mb{S}_a \times \mb{S}_{a+e_{x}})\cdot \hat{x} +(\mb{S}_a \times \mb{S}_{a+e_{y}})\cdot \hat{y} \right] \,.
\label{DiscreteModel}
\end{align}
Here, $\textbf{S}_a=(S^x_a,~S^y_a,~S^z_a)$ is the spin of size $S$ at lattice site $a$ and ${\bf e}=(e_x,e_y)$ is the lattice unit vector. 
We assume  the on-site anisotropy constant to be positive, $\tilde K>0$, so that there is an easy axis along $z$ and fix the chirality of the Dzyaloshinskii-Moriya (DM) interaction, $\tilde D$. The strength of the magnetic field along the $z$ axis is $\tilde H$ and the exchange interaction constant is assumed to be ferromagnetic, $\tilde J>0$. 

Using coherent state representation for the spins (assuming $S\gg 1$) and passing to the continuum limit~\cite{BraunPR96}, we find that Eq.~(\ref{DiscreteModel}) corresponds to the free energy of the form
\begin{align}
\mathcal F(\mb{m})= J \sum_{i=x,y} \left( \frac{\pt \mb{m}}{\pt \tilde r_i}\right)^2 +D \mb{m} \cdot \nabla \times \mb{m}  - K m_z^2  - H m_z    \,,
\label{FreeEnergy}
\end{align}
\begin{figure}[b!]
  \includegraphics[scale=0.25]{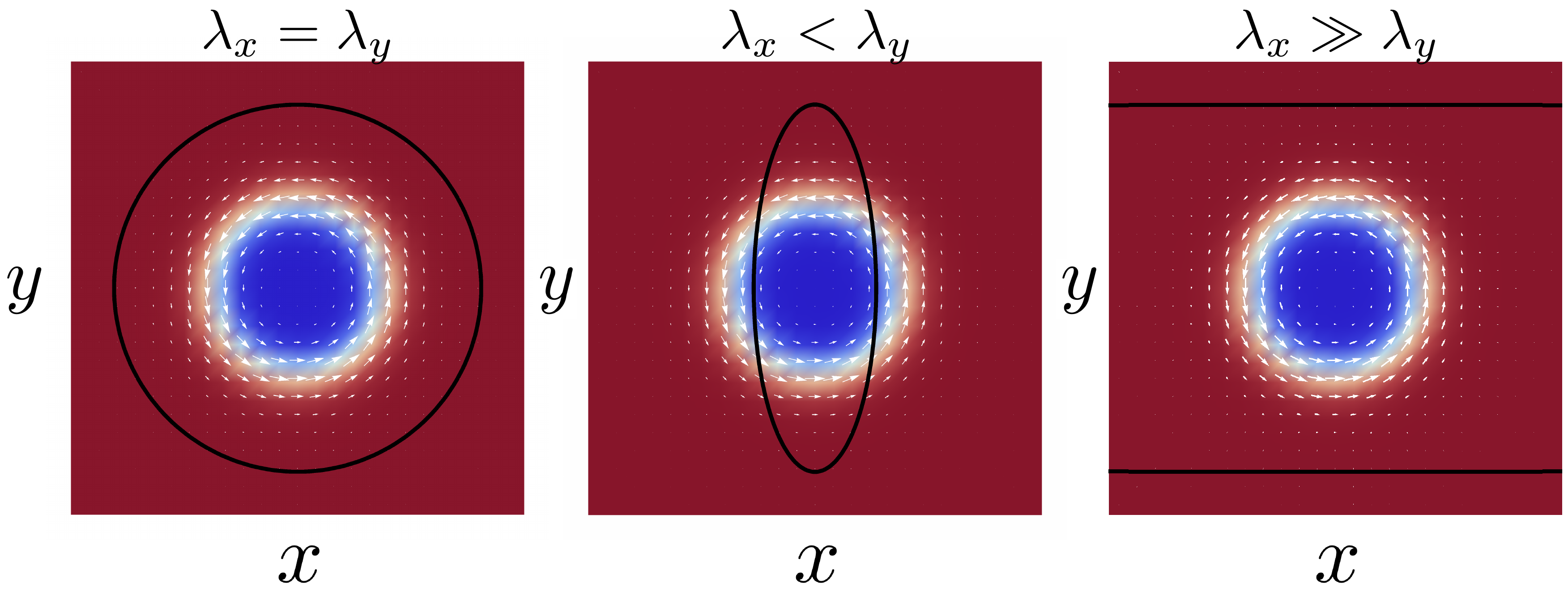}
  \caption{\ Pictorial representation of a skyrmion in the presence of a magnetic trap described by Eq.~(\ref{MagTrap}) for different magnetic trap sizes. The colored surface represents the out-of-plane component of the magnetization texture of a skyrmion with positive maximum at the center (blue), while black solid line represents the size of the trap.}
   \label{fig:MagneticTrap}
\end{figure}
where $J=\tilde{J} S^2$, $K= \tilde{K} S^2/\alpha^2$, $H=g \mu_B \tilde{H} S^2/\alpha^2$, and $D= \tilde{D} S^2 /\alpha$. We now introduce dimensionless spatial variables $\mb{r}=\tilde{\mb{r}}/(l \alpha)$ where $l=\tilde{J}/ \tilde{D}=J/ (D\alpha)$, while energies are measured in units of $\eL=  \tilde{J}S^2=J$. The energy density in reduced units is defined as $w(\Phi,\Pi)= \frac{(l\alpha)^2}{\eL} \mc{F}(\Phi,\Pi)$ (see Eq.~\eqref{EnergyDensity}).
We consider the regime of isolated skyrmions, which exist for a wide range of parameters $K$ and $H$ as a stable or metastable state. A skyrmion with $Q=-1$ is parametrized by (see Appendix \ref{sec:ap1})
\begin{equation}
\Theta_0(\mb{r})=\Theta_0(\rho) \,,\quad \Phi_0(\mb{r})= \phi \pm \pi/2 \,,
\label{Solution}
\end{equation}
while the helicity, defined by the sign in $\Phi_0$, is commensurate with the DM interaction. The structure of the stationary skyrmion $\Theta_0(\rho)$ is determined by the Euler-Lagrange equation derived from skyrmion energy $w(\Phi_0, \Pi_0)$ and is given in Eq.~\eqref{EulerEq}. Because there is no known analytic solution for $\Theta_0(\rho)$, the following function can be used as an approximate solution,
\begin{align}
\Theta_0(\rho)=2 \tan^{-1} \left( \frac{\lambda}{\rho}e^{-\frac{\rho -\lambda}{\rho_0}} \right)\,.
\label{Profile_Sm}
\end{align}
Roughly speaking, $\rho_0=\sqrt{2/(2\kappa +h)}$ determines the amount of spins that are \textit{noncollinear} with the easy axis, while $\lambda$, which we obtain numerically using an algorithm based on Runge-Kutta formulas \cite{Algorithm}, is the skyrmion radius of the area for which the spins are \textit{parallel} to the easy axis. Here, $\kappa=\tilde K \tilde J/\tilde D^2$ and $h=g\mu_B\tilde H \tilde J/\tilde D^2$ are the dimensionless parameters describing easy axis anisotropy and magnetic field, respectively (see Table.~\ref{Units}). In the opposite limit of a large-radius skyrmion the magnetization profile is described by
\begin{equation}
\cos \Theta_0(\rho) = \tanh (\frac{\rho - \lambda}{\Delta_0}) \,,
\label{Profile_Lrg1}
\end{equation} 
where the parameters $\lambda$ and $\Delta_0$ are calculated numerically by fitting the approximate function \eqref{Profile_Lrg1} to the numerical solution of the Euler-Lagrange equation \eqref{EulerEq}. Magnetization profiles of skyrmions in the small- and large- radius limit obtained numerically are depicted in Fig.~\ref{Profile}. A detailed discussion of the magnon eigenstates $\Psi_n$ is provided in Appendix \ref{sec:ap2}. Apart from the magnon scattering states $\Psc_n$ that lie above the magnon gap $\varepsilon_{\mbox{\tiny{MS}}}=\kappa+ h/2$, induced by the anisotropy and magnetic field, there exist massive internal modes $\Pbs_n$ that are found for energies $0 < \varepsilon_n \leq \varepsilon_{\mbox{\tiny{MS}}}$. These bound states correspond to deformations of the skyrmion into breathing modes and were discovered numerically in Refs.~\onlinecite{LinPRB14,SchuttePRB14}.
\begin{figure}[t!]
\includegraphics[width=1\linewidth]{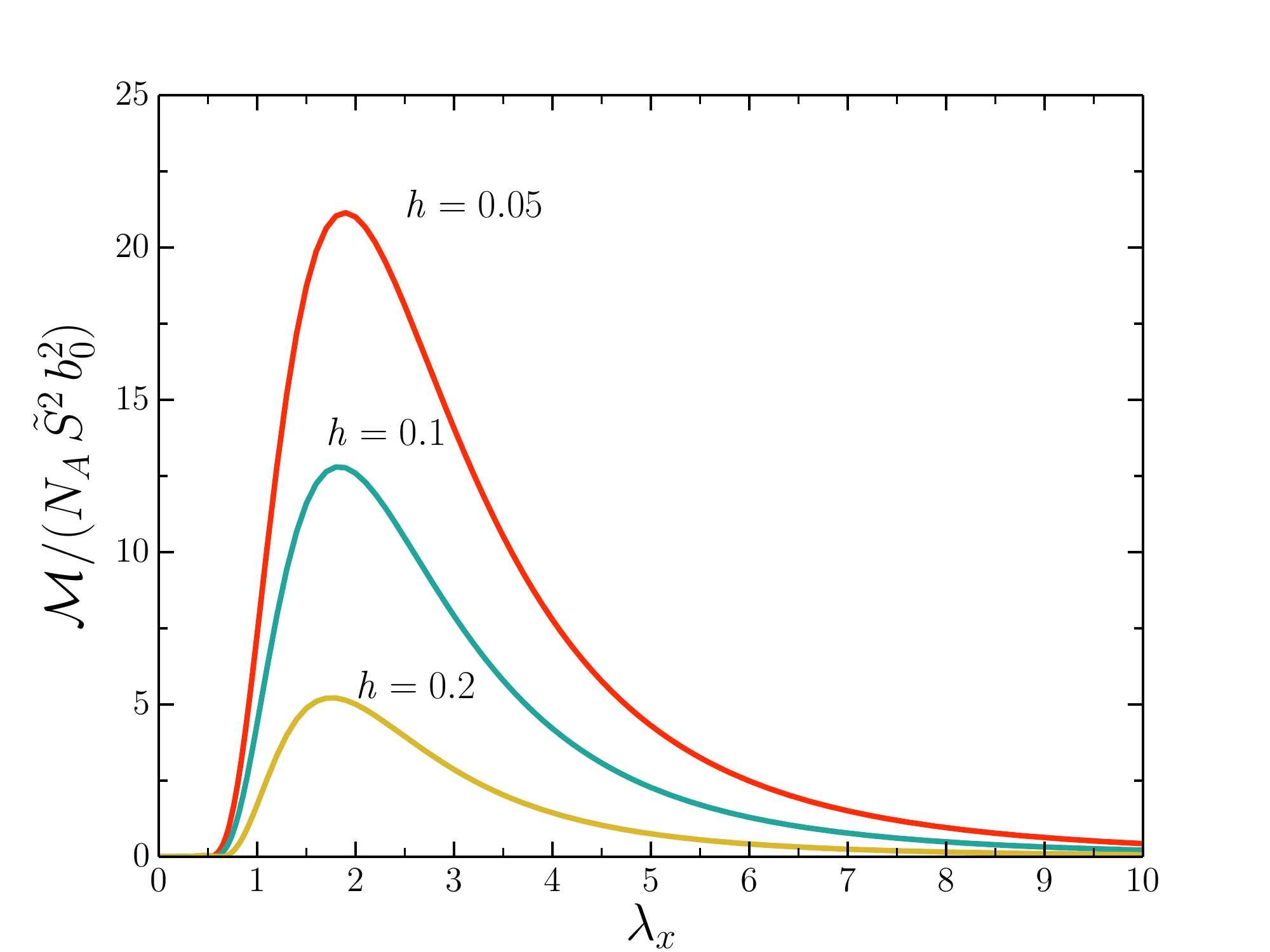}
 \caption{\ Dependence of the mass term $\mc{M}$ on the trap size $\lambda_x=\lambda_y$ of an isotropic magnetic trap of strength $b_0=0.01$ at $T=0$. The skyrmion configuration is described by the parameters $\kappa=0.7$, and three values of magnetic field $h= 0.2,~0.1,~0.05$ with skyrmion radius $\lambda=1.26,~1.57,~1.79$, respectively.}
 \label{MassIs}
\end{figure}

\section{Induced Skyrmion Mass}
\label{subsec:Mass Zero T}
The mass tensor $\mc{M}_{ij}^{0}$ \eqref{Mass0} of a magnetic skyrmion proportional to the effective spin $N_A \bar{S}$ is now investigated, which is nonzero only in the presence of a perturbation in the free energy that breaks translational symmetry. We should stress that once a perturbation is considered, the skyrmion acquires a mass equal to $\mc{M}^0_{ij}$ at any  temperature. We consider three isotropic mechanisms for the generation of a skyrmion mass, all breaking translational invariance. (1) a magnetic trap, generated by a nonuniform magnetic field, (2) a local defect that alters the exchange constant, and (3) a periodic variation in the exchange constant. In all three cases, we assume the translationally noninvariant terms to be much smaller than those that are translationally symmetric, $|\int d \textbf r \mathcal V(\Phi_0,\Pi_0)|\ll S_{E}$, and they can thus be treated perturbatively. In all cases where an isotropic potential in $x$ and $y$ direction is applied, we find $\mc{M}^0_{xx} = \mc{M}^0_{yy}= \mc{M}$.
\begin{figure}[t!]
\includegraphics[width=1\linewidth]{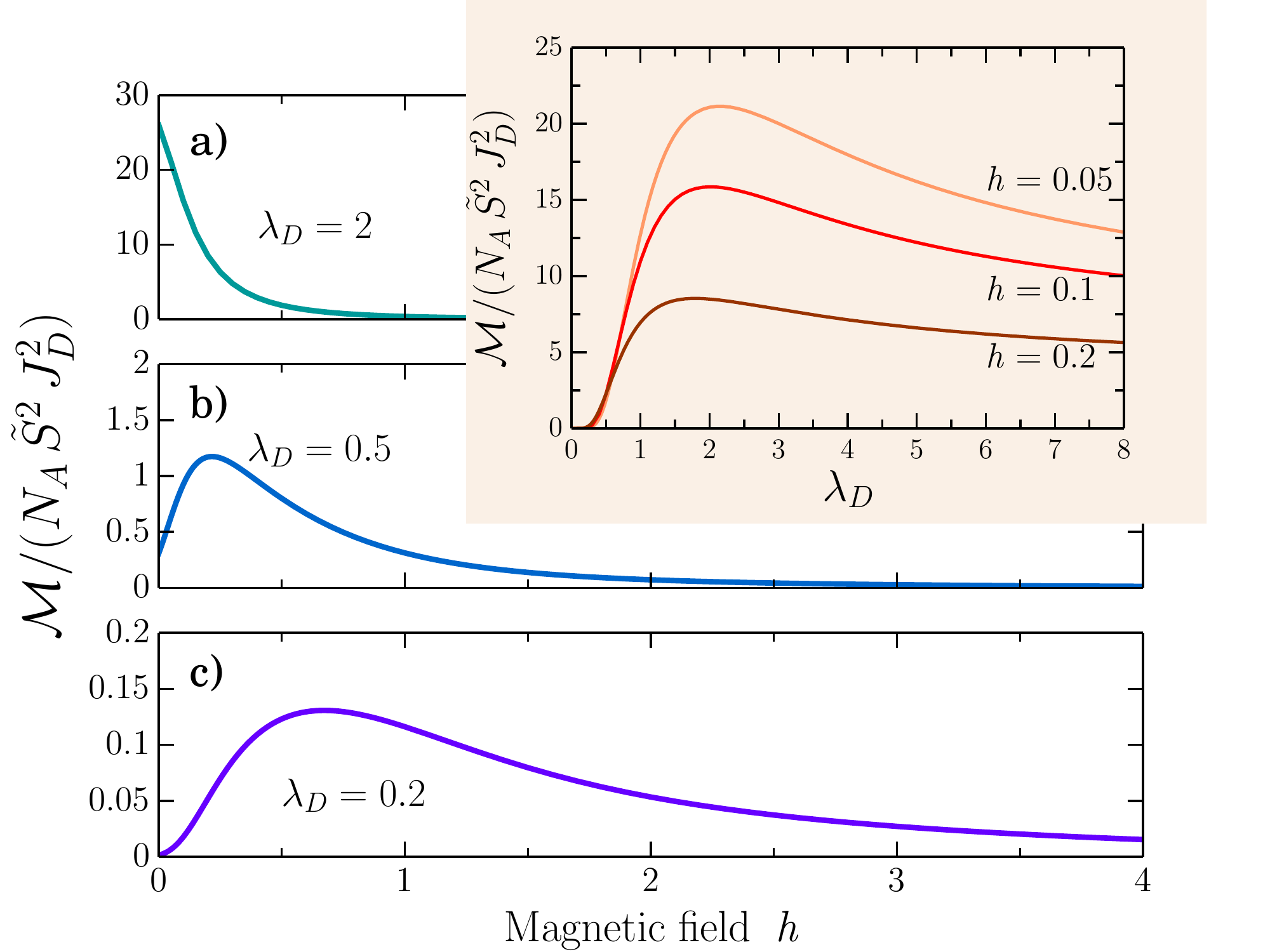}
 \caption{\ Magnetic field dependence of the mass term $\mc{M}$ in the presence of a defect for a skyrmion with $\kappa=0.7$, $J_D=0.1$ at $T=0$. We depict three different values of defect size $\lambda_D=2$ (a), $0.5$ (b), and $0.2$ (c). The inset shows the dependence of $\mc{M}$ on the defect size $\lambda_D$ for a skyrmion with $h= 0.2,~0.1,~0.05$ and skyrmion radius $\lambda=1.26,~1.57,~1.79$, respectively. }
\label{massLD}
\end{figure}

A simple realization of a magnetic trap is 
\begin{equation}
\mb{b}(\mb{r})= - b_0 e^{- \frac{x^2}{\lambda_x^2} -\frac{y^2}{\lambda_y^2}} ~\hat{z} \,,
\label{MagTrap}
\end{equation}
where $\lambda_x$ and $\lambda_y$ characterize the size of the trap along the $x$ and $y$ directions, respectively. See Fig.~\ref{fig:MagneticTrap} for a visualization of the radially symmetric, $\lambda_x=\lambda_y$,  and asymmetric, $\lambda_x\neq\lambda_y$, magnetic traps. Because the propagating magnon modes are suppressed, we numerically calculate the mass terms by including only the lowest energy bound states, which are themselves found variationally, in the sum of Eq.~(\ref{MassTerm}). In Fig.~\ref{MassIs} we plot these results as a function of $\lambda_x$ for a radially symmetric trap, $\lambda_y=\lambda_x$, where we have a chosen $\kappa=0.7$, $b_0=0.01$ and three values of magnetic field $h= 0.2,~0.1,~0.05$ with skyrmion radius $\lambda=1.26,~1.57,~1.79$ respectively. For $\lambda_x$ smaller than a critical value $\lambda_x^{cr}\sim \lambda$, the mass increases with increasing $\lambda_x$, while for $\lambda_x\gtrsim\lambda_x^{cr}$ the mass decreases monotonically. 
\begin{figure}[t!]
\centering
\includegraphics[width=1\linewidth]{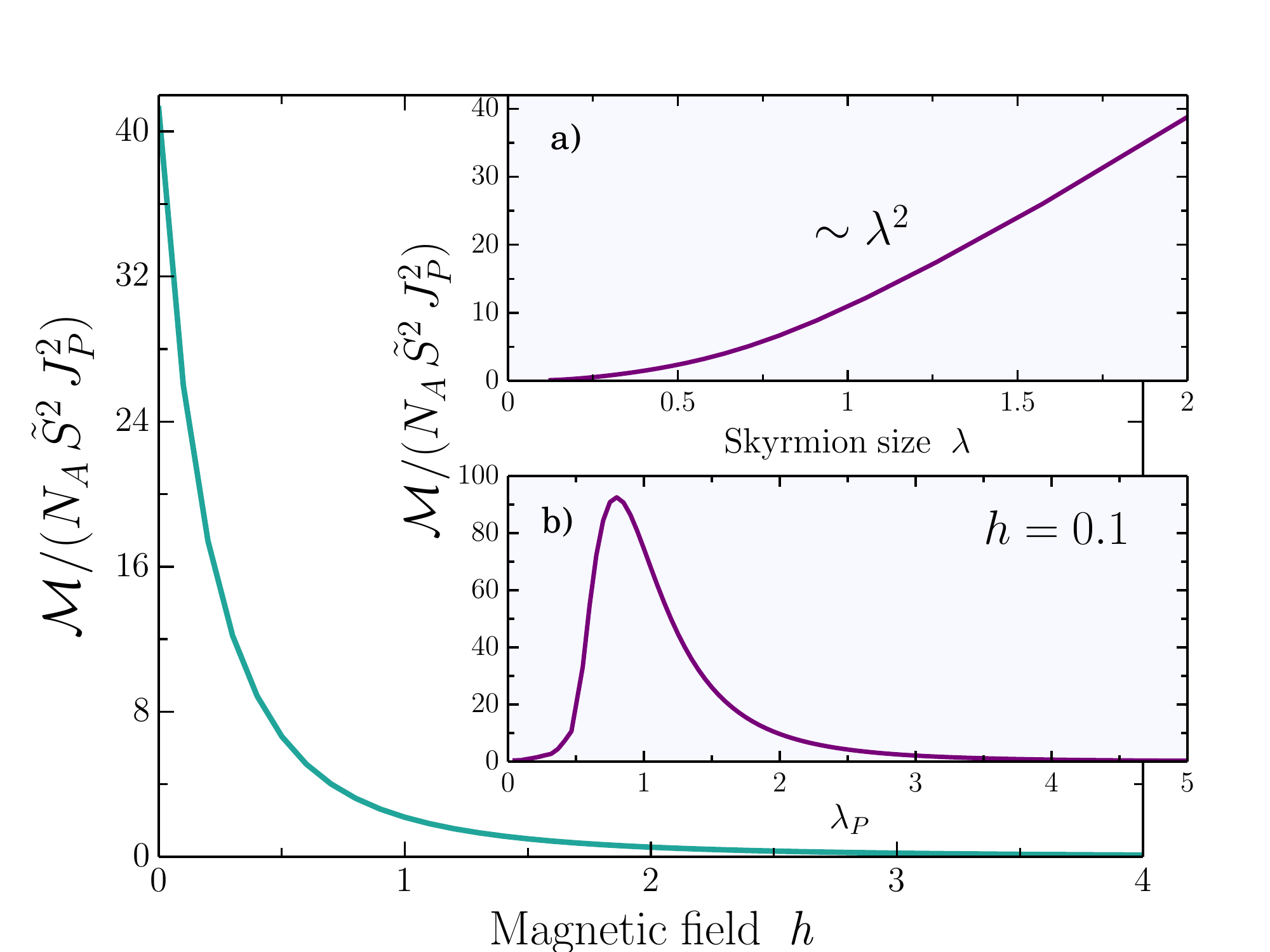}
\caption{\  Magnetic field dependence of the mass term $\mc{M}$ for spatially dependent periodic exchange interactions with $\lambda_P=1.5$, $J_P=0.1$ and $T=0$. Inset a) depicts the $\lambda^2$ dependence of $\mc{M}$, while inset b) illustrates $\mc{M}$ as a function of period $\lambda_P$ for $h=0.1$. }\label{MassPh}
\end{figure}

When there is a local defect in the crystal structure of the chiral magnet, e.g. at the origin, the exchange interaction is a function of position around that point which we model as
\begin{equation}
J\rightarrow J_0-J_D e^{-\rho /\lambda_D} \,,
\label{ExchDef}
\end{equation} 
where $J_D$ and $\lambda_D$ parametrize the strength and size of the defect, respectively.

In Fig.~\ref{massLD} we plot the skyrmion mass due to such a defect as a function of applied magnetic field for several values of $\lambda_D$ and $J_D=0.1$. The behavior of Fig.~\ref{massLD} implies that as long as $\lambda_D < \lambda$,  the mass $\mc{M}$ increases with increasing (decreasing) magnetic field (Skyrmion size), peaking around $\lambda_D \sim \lambda$ and then for $\lambda_D > \lambda$ decreases with decreasing skyrmion size (see Fig.~\ref{massLD}(b) and (c)). We see no peak when $\lambda_D=2$, Fig.\ref{massLD}~(a), because the skyrmion radius for all positive magnetic fields is smaller than the size of the defect. In addition, the dependence of $\mc{M}$ on $\lambda_D$ is presented in the inset of Fig.~\ref{massLD}, where a peaked behavior is suggested with a maximum around $\lambda$. We note that although it has been shown that a model with spatially dependent exchange interactions similar to Eq.~\eqref{ExchDef} acts as a pinning potential\cite{ShiZengLi}, {\it i.e.}, $\partial\mathcal V/\partial \mathrm{R}_i$ in Eq.~\ref{sEoM}, our findings  show that additionally a finite mass is generated. Finally, because of the discrete underlying lattice the exchange interaction can obtain a periodic modulation along the lattice vectors,
\begin{equation}
J \rightarrow J_0 -J_P \cos( x/\lambda_P) -J_P \cos(y/\lambda_P) \,,
\end{equation} 
where $J_P$ and $\lambda_P$ are the strength and period, respectively, by which the exchange deviates from $J_0$. In Fig.~\ref{MassPh} we plot the magnetic field dependence of $\mc{M}$ for $\lambda_P =1.5$ and $J_P=0.1$. Focusing on the dependence of $\mc{M}$ on the skyrmion radius $\lambda$, illustrated in the inset of Fig.~\ref{MassPh} (a), it becomes apparent that $\mc{M}$ grows with the area $\lambda^2$. Further results are provided in the inset of Fig.~\ref{MassPh}~(b), where we examine the dependence of the mass on the period $\lambda_P$ for $h=0.1$, $\lambda=1.57$, and $J_P=0.01$. Figure~\ref{MassPh}~(b) reveals a peaked behavior around $\lambda/2$.

\subsection{ Massive Skyrmion in quasi-1D confinement}
\begin{figure}[t!]
\centering
\includegraphics[width=1\linewidth]{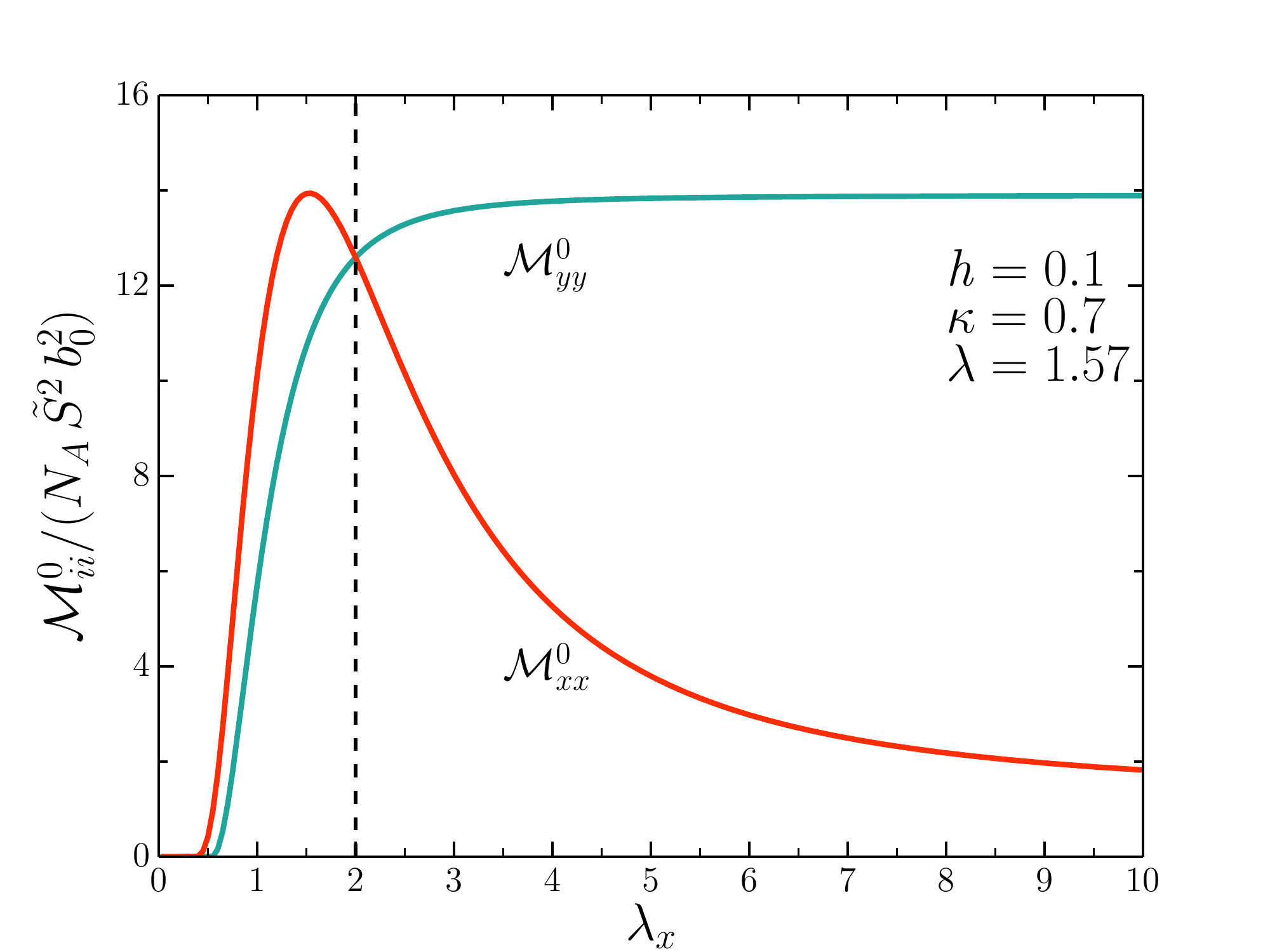}
\caption{\  Dependence of the mass terms $\mc{M}^0_{xx}$ and $\mc{M}^0_{yy}$ on the trap size $\lambda_x$ of an anisotropic magnetic trap of strength $b_0=0.01$ and fixed $\lambda_y=2$. The skyrmion configuration is described by the parameters $\kappa=0.7$, $h=0.1$, and $\lambda=1.57$. The dashed vertical line indicates the location of $\lambda_y=2$.}\label{MassAnis}
\end{figure}

The case of a magnetic skyrmion confined in a quasi-one-dimensional space can be realized by applying an anisotropic magnetic trap whose size  is much larger in one spatial direction than in the other, {\it i.e.}, $\lambda_x \gg \lambda_y$. A pictorial representation of a skyrmion in the presence of an anisotropic magnetic trap is provided in Fig.~\ref{fig:MagneticTrap}. When $\lambda_x \ne \lambda_y$, the elements of the mass tensor $\mc{M}_{ij}^0$ are no longer equal, but acquire different values depending on the size of the magnetic trap. In Fig.~\ref{MassAnis} we plot the dependence of $\mc{M}^0_{xx}$ and $\mc{M}^0_{yy}$ on $\lambda_x$ for a fixed value of $\lambda_y=2$ and a skyrmion with $\kappa=0.7$, $h=0.1$ and $\lambda =1.57$. We find that $\mc{M}^0_{xx}$ is  maximized at $\lambda$; conversely, $\mc{M}^0_{yy}$ grows until $\lambda_y\sim\lambda_x$ upon which it approaches a constant value. 

A corollary of the preceding discussion is that the highly anisotropic magnetic trap acts as a {\it magnetic quantum wire}, along which the skyrmion moves as a massive particle due to its one-dimensional confinement. Our investigation demonstrates that skyrmions observed in confined geometries such as magnetic nanowires \cite{Mehlin15} exhibit fundamentally different dynamical behavior compared to the one in unconfined 2D geometries. It would be interesting to test this prediction.
\section{Skyrmion Mass at Finite Temperature}
\label{subsec:Mass Finite T}

\begin{figure}
\centering
\includegraphics[width=1\linewidth]{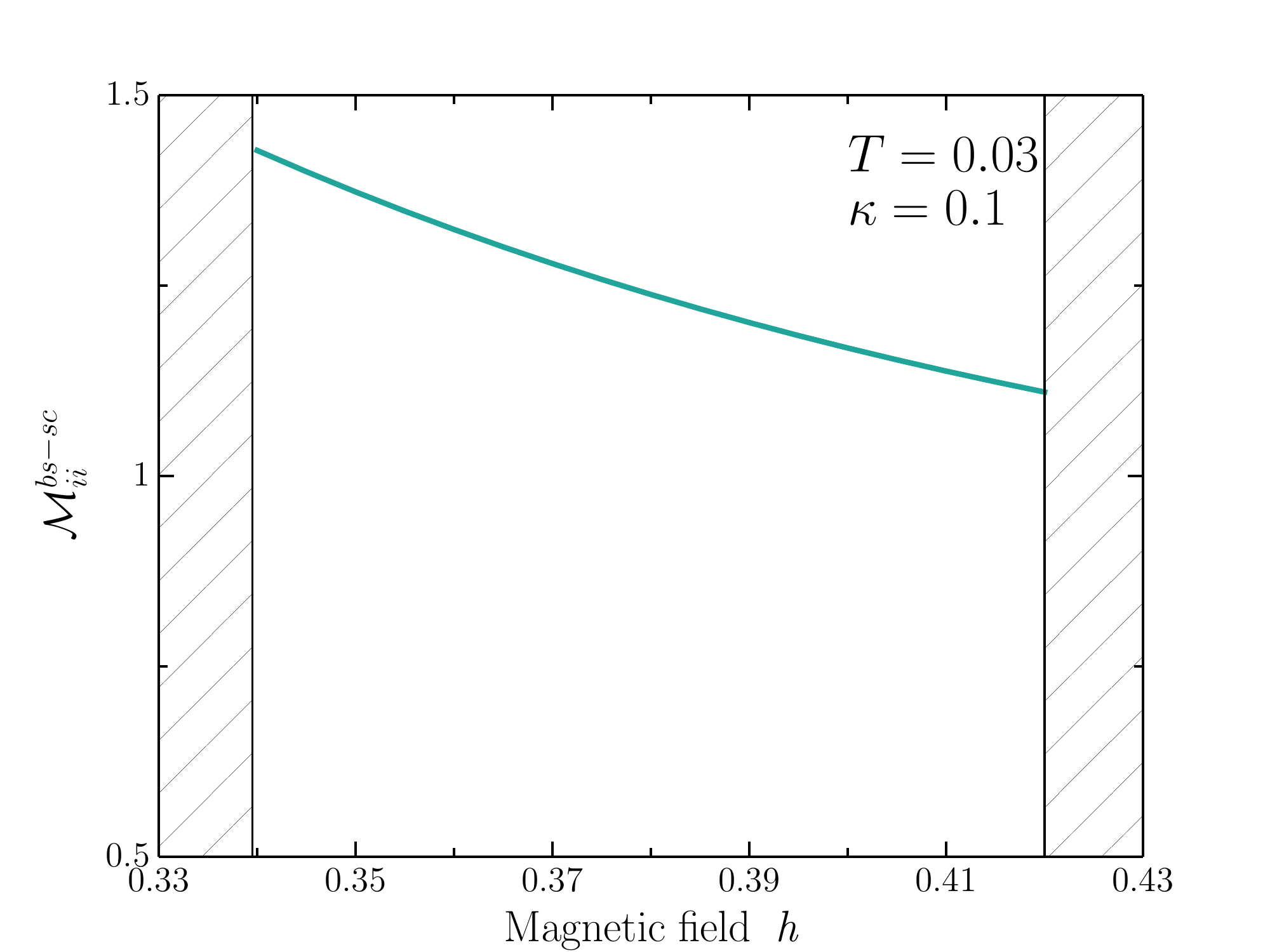}
	\caption{\ Magnetic field dependence of diagonal mass tensor elements $\mathcal{M}_{ii}^{bs-sc}$ for a skyrmion with $\kappa=0.1$, at temperature $T=0.05$ and for $S=1$ and $\tilde{J}/\tilde{D}=4$. The magnetic field region where bound states contribute to the mass is bounded by the critical field below which the skyrmion becomes unstable (left vertical line) up to the field where the bound state energy passes to the continuous spectrum (right vertical line).}\label{MassBSC}
\end{figure}
When no translational symmetry breaking term is present, the only contribution to the mass is given by the term $\mc{M}_{ij}^{T}$, which is independent of $N_A\bar{S}$ and finite even at vanishingly small temperatures. The series expression Eq.~\eqref{MassT} requires the summation over the full magnon spectrum, that is over scattering states $\Psc_{m,k}$ which are classified by the azimuthal quantum number $m$ as well as the radial momentum $k \geq 0$ with energy $\varepsilon_k=\varepsilon_{\mbox{\tiny{MS}}} +k^2$, with $\varepsilon_{\mbox{\tiny{MS}}}=\kappa+h/2$. In addition to scattering states, one needs to take into account localized modes $\Pbs_m$, classified by $m$ with energies in the range $0 < \varepsilon_m \leq \ems$. Therefore we consider the following terms,
\begin{align}
\mc{M}_{ij}^{T}= \Msc + \Mbs +\Mbsc \,,
\end{align}
where $\Msc$ denotes the contribution from scattering states, $\Mbs$ the contribution from bound states,  and $\Mbsc$ takes into account combinations of bound-scattering states. Here, we focus on large-radius skyrmions for low magnetic fields and easy axis anisotropy, where the skyrmion is expected to support more internal modes with energy below $\varepsilon_{\mbox{\tiny{MS}}}$ than for small-radius skyrmions. As the magnetic field or easy-axis anisotropy is increased, the radius decreases and the modes sequentially leave the gap region, pass to the continuous spectrum and transform into quasilocalized modes. Therefore, there is a bounded parameter region of $h$ and $\kappa$ where the requirement $\varepsilon_m \leq \ems$ is fulfilled.

A calculation of bound and scattering states is presented in Appendix~\ref{sec:ap2}, while explicit expressions for $\Msc$, $\Mbs$, and $\Mbsc$ are given in Appendix~\ref{sec:ap3}. From the form of the matrix elements $\mc{B}_{ij}^{ns,n's'}$ of Eq.~\eqref{ElementB} we conclude that only combinations of states with angular momentum difference $\Delta m = \pm 1$ contribute to the finite $T$ skyrmion mass $\mc{M}_{ij}^{T}$. From the variational calculation of bound state energies in the large skyrmion limit for the parameter range considered here, we find two localized modes below the magnon continuum with $m=0$ and $m=2$. Therefore, we find vanishing contributions from the term $\Mbs$ because the zero mode with $m=1$ is excluded from the summations. The behavior of $\Mbsc$ for a skyrmion with $\kappa=0.1$, $S=1$ and $\tilde{J}/\tilde{D}=4$ as a function of $h$ for $T=0.05$ is summarized in Fig.~\ref{MassBSC}, while the main features of $\Mscx$ as a function of $T$ and $h$ are depicted  in Fig.~\ref{QMass}. Explicit expressions on summations over scattering states are given in Appendix~\ref{sec:ap3}. Summations over the quantum number $m$ converge rapidly and are bounded between $-7 \leqslant m  \leqslant7$. In Fig.~\ref{QMass}, we calculate $\mc{M}^{T}_{ij}$ for a discrete set of $h$ and $T$ and then we adopt a process to obtain the mass as a smooth function of $(h,T)$.

We note that Eq.~\eqref{MassTerm} for the mass is  valid up to temperatures of the order of the magnon gap,  \textit{i.e.}, $T \lesssim \egap= \frac{2 \gap }{S (\tilde{J}/\tilde{D})^2}$.  For Fig.~\ref{QMass} we have $\gap=\ems$. Choosing $\tilde{J} = 32 \times 10^{-4}$ eV, $\tilde{D} = 8 \times 10^{-4}$ eV,  $\alpha = 9$ \AA ~, $\tilde{K} =0.2 \times 10^{-4}$ eV ($\kappa = 0.1$), $S=1$, and for a skyrmion stabilized at $\tilde{H} = 864 $ mT ($h=0.5$), the effective mass $\Msc$ has a temperature dependence given in Eq.~\eqref{MassT} which is valid up to $\tilde{T} \lesssim 1.62$ K. Similarly, for a skyrmion stabilized at $\tilde{H} = 518$ mT ($h=0.3$), the highest temperature for which Eq.~\eqref{MassT} is reliable is $\tilde{T} \lesssim 1.16 $ K. The  mass in physical units is then given by $\Msc \cdot 0.7 \times 10^{-28}$ kg. 
\begin{figure}[t]
\centering
\includegraphics[width=1\linewidth]{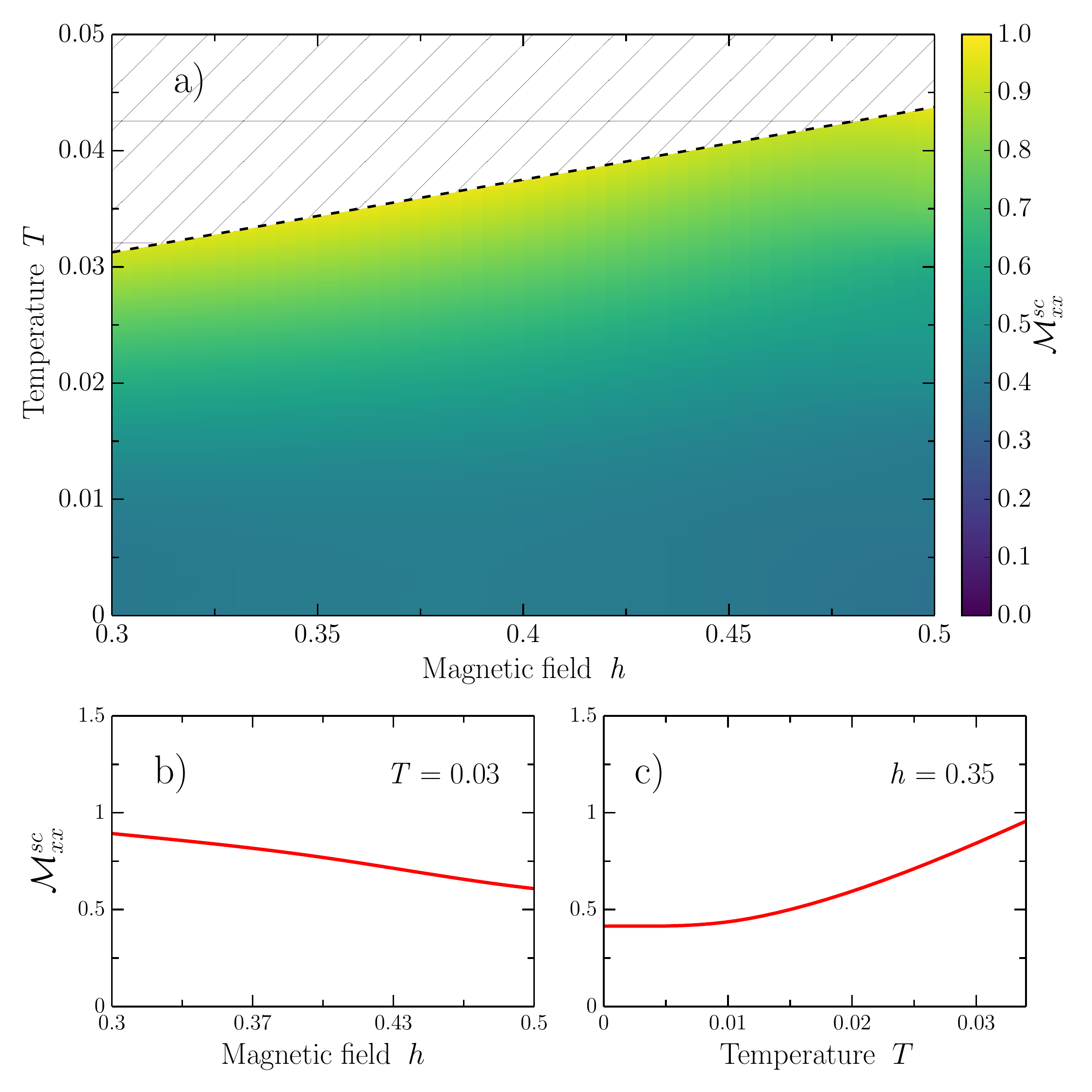}
	\caption{\ a) The colored surface represents the skyrmion mass $\mathcal{M}_{xx}^{sc}$ in the quantum regime calculated from Eq.~\eqref{MassTerm} for $\kappa=0.1$, $S=1$, and $\tilde{J}/\tilde{D}=4$ as a function of dimensionless magnetic field $h$ and  temperature $T$ (for units see text). The dashed line $T= \frac{2 (\kappa+ h/2)}{S (\tilde{J}/\tilde{D})^2} $ indicates 
	the range of validity of  Eq.~\eqref{MassTerm}.
	b) Magnetic field dependence of $\mathcal{M}_{xx}^{sc}$ for a fixed temperature $T=0.03$ and c) temperature dependence of $\mathcal{M}_{xx}^{sc}$ for a fixed magnetic field $h=0.35$.}\label{QMass}
\end{figure}

\section{Discussion}
\label{sec:Discussion}
Equation \eqref{MassTerm} is the first closed  formula for the effective mass of a skyrmion, obtained microscopically in the presence of arbitrary external perturbations arising from defects, nonuniform magnetic fields, and external potentials and at finite temperature. Skyrmion phases have been observed in several metallic ferromagnets, including MnSi \cite{Muhlbauer09} and FeGe \cite{SkyrmionFeGe}, as well as in the insulator Cu$_2$OSeO$_3$ \cite{Seki12,Adams12}. Typical parameter values for Cu$_2$OSeO$_3$ crystals are $\tilde{J} = 33.36 \times 10^{-4}$ eV, $\tilde{D} = 7.47 \times 10^{-4}$ eV and $\alpha=8.911$ \AA ~($\tilde{J}/\tilde{D}=4.46$) \cite{Janson14}. The induced mass in physical units is then given by $\mc{M}_{ij}^{0} \times N_A \cdot 4 \cdot 10^{-26}$ kg, which corresponds to $5 \times 10^4$ electron masses. The scale that separates the  quantum from the finite-temperature regime is given by the magnon gap $\tegap \sim 2$ K. A typical scale for the quantum mass $\mc{M}_{ij}^{T}$ is $\sim 10^{-28}$ kg with a temperature dependence given in Eq.~\eqref{MassT} and plotted in Fig.~\ref{QMass}. 
For high temperatures, $k_B\tilde{T} > \tegap$, our approximation breaks down since the reduction of the  damping kernel to an effective mass in Eq.~\eqref{MassTerm} is not valid anymore.
We remark that the skyrmion dynamics in this classical regime, $\tegap < k_B \tilde{T} \ll J$, was studied in Ref.~\cite{SchuttePRB90} 
via a stochastic LLG equation giving a mass that is independent of temperature. This suggests that the weak temperature dependence of the mass found here in the quantum regime is flattened off 
by some second-order processes activated 
at higher temperatures, such as magnon-magnon interactions (neglected in this work). It would be interesting to include such higher-order terms in our microscopic approach systematically. We leave such a study for future work.

Further, we note that the authors of Ref.~\cite{SchuttePRB90} predict a strong frequency dependence of the dynamics of the skyrmion motion, as a result of the presence of time-dependent forces, as well as the coupling of the skyrmion to magnon excitations. Among the results so obtained is a large numerical value temperature independent effective mass $m(\omega)$ as well as a frequency-dependent Gilbert damping $ D(\omega)$. To make contact with these results, a frequency-dependent mass is identified as $m(\omega) \rightarrow 2~ \mbox{Re} \gamma_{ii}^{\omega_\nu}$ and Gilbert damping is $D(\omega) \rightarrow -2 \omega_{\nu}  ~\mbox{Im}\gamma_{ii}^{\omega_\nu}$, where $\gamma_{ii}^{\omega_\nu}$ due to magnons is known exactly and  given in Eq.\eqref{DampFreq}. Since eigenenergies and frequencies scale as $1/\lambda^2$, where $\lambda$ is the skyrmion radius, from our explicit expressions we obtain that, $m(\omega) \sim \lambda^2$ (see inset of Fig.~\ref{MassPh}) while $D(\omega)$ is independent of $\lambda$, a result which is consistent with Ref.~[\onlinecite{SchuttePRB90}]. Moreover, a large inertial mass is found for the case of thermal diffusion which induces a large skyrmion distortion (strong translational symmetry breaking term), while an almost vanishing mass was found for the case of an electric current-driven motion, when the skyrmion flows with the current with little distortion (weak translational symmetry breaking term). This result is consistent with the interpretation of skyrmion mass generation provided by our calculation; any term which explicitly breaks translational symmetry will induce a finite mass even at vanishingly small temperature. 

Quite generally, the skyrmion dynamics obtained by the LLG equation and the (quantum) damping obtained exactly in closed form in the present work are complementary methods. The former provides a reliable framework especially when a microscopic description is difficult to obtain, for example the inclusion of a large number of microscopic degrees of freedom such as electrons. However, the latter can provide a detailed microscopic origin of dissipation beyond the classical limit and beyond the predictability of the LLG equation. 

In Ref.~[\onlinecite{Iwasaki}] an inertial term was estimated that appears due to a confining potential, independent of the skyrmion size $\lambda$ and proportional to the stiffness of the harmonic confining potential. Finally, it is also worth mentioning that an approximate evaluation of the mass term of a skyrmion magnetic bubble, a circular domain wall stabilized by long-range interactions has been found in Ref.~[\onlinecite{Makhfudz}], derived by generalizing Thiele's approach beyond steady motion. We emphasize that the mass tensor depends on the inverse of the magnon energy $\varepsilon_n$. We therefore expect that bound states, which correspond to deformations of the skyrmion with energies below the gap, have a larger contribution. This finding is consistent with Ref.~[\onlinecite{Makhfudz}] where an approximate value for the mass of a magnetic bubble domain has been calculated by considering contributions from distortions of the circular domain wall. In addition, a large value of inertia mass of a magnetic bubble has been experimentally reported in Ref.~[\onlinecite{Buttner15}]. The bubbles were subjected to a magnetic disk and the presence of an inertial mass was attributed to a breathing mode associated with a change of the bubble size. 

While several studies demonstrate the association of inertia mass to deformations of the skyrmion, an explicit analytical formula has been missing so far and is provided here. Our detailed calculations shed light on the mechanism of mass generation, and we also provide an analytic formula, valid in the presence of arbitrary external perturbations which can serve as a basis for a variety of further studies.

\section{Concluding Remarks}
\label{sec:Conclusions}
\par

The main task of the paper is to describe dissipation of a magnetic skyrmion using the canonical scheme of quantization. Within this formalism the skyrmion position is promoted to a time-dependent dynamical variable and a finite perturbation theory in terms of fluctuations around the skyrmionic configuration is performed. The interaction of the skyrmion with these magnon modes gives rise to a damping term that is found to be time dependent, in contrast to the typical Gilbert damping. The time-dependence of damping kernel indicates that damping depends on the velocity of the skyrmion of the past. 

Perhaps, the most interesting feature is that the effect of damping is reduced to a mass term in some limits. We demonstrate that a massless skyrmion at the classical level is a consequence of the translational symmetry assumed for the system. Our investigation suggests that, even at vanishingly small temperature, the skyrmion mass is nonzero in the presence of an external perturbation arising from defects, nonuniform magnetic fields and external potentials. Among the translational symmetry breaking terms we consider here, we emphasize the case of a quasi-one dimensional wire provided by an anisotropic magnetic trap, along which the skyrmion moves as a massive particle owing to its confinement. To complete the description we also examine damping in the absence of perturbations, and we calculate an effective spin-independent quantum term with an explicit temperature dependence and a nonvanishing particle-antiparticle contribution at zero temperature. This result reflects the fact that not only thermal but also quantum fluctuations  contribute to a mass term.

This picture suggests that in the absence of perturbations, massive magnon modes are activated at finite temperatures. A utilization of the improved skyrmion dynamics calculated in the present paper could be towards the direction of investigating the possibility of depinning via quantum tunneling, a behavior that has been studied both theoretically \cite{Braun97} and experimentally \cite{Brooke01} in similar mesoscopic magnetic structures, such as magnetic domain walls. 

In our present study, we exclusively focus on damping caused by intrinsic mechanisms originating from the interaction of the skyrmion with magnon modes which it generates and couples back to, giving rise to damping. Additional sources of damping could arise if one takes into account time-dependent forces or the interaction of the skyrmion to extrinsic degrees of freedom, like a current of electrons or phonons. However, this is beyond the scope of this work, and we leave it as a motivation for further studies making use of the general formalism developed here.

\section*{ACKNOWLEDGEMENTS}
We would like to acknowledge Peter Stano for useful discussions. This work was supported by the Swiss National Science Foundation  and NCCR QSIT.
\appendix
~~
\section{Skyrmion as Saddle Point Configuration}
\label{sec:ap1}

In this Appendix we give the explicit form of the skyrmion by finding the approximate saddle point configuration of the free energy of Eq.~\eqref{FreeEnergy}. It is convenient to introduce dimensionless spatial variables $\mb{r}=\tilde{\mb{r}}/(l \alpha)$, where $\l= \tilde J/\tilde D= J/(D \alpha)$, and $\eL=J= \tilde J S^2$ is a characteristic energy scale. The energy density of the spin configuration in reduced units $w(\Phi,\Pi)= \frac{(l \alpha)^2}{\eL} \mc{F}(\Phi,\Pi)$ in polar cooordinates $\mb{r}=(\rho \cos \phi,\rho \sin \phi)$ is given by
\begin{widetext}
\begin{eqnarray}
w(\Phi , \Pi)&=&  (\nabla \Theta)^2+ \sin^2\Theta( \nabla \Phi)^2 - \kappa \cos^2\Theta - h \cos\Theta  \nonumber \\
&+& \left(\cos(\phi-\Phi)\frac{\pt \Theta}{\pt \rho} - \sin\Theta\cos\Theta \sin(\phi-\Phi)\frac{\pt \Phi}{\pt \rho} - \frac{1}{\rho} \sin(\phi -\Phi)\frac{\pt \Theta}{\pt \phi} -\frac{1}{\rho}\sin\Theta\cos\Theta\cos(\phi-\Phi)\frac{\pt \Phi}{\pt \phi} \right)\,,~~~~
\label{EnergyDensity}
\end{eqnarray}
\end{widetext}
where $\kappa=KJ/D^2$ and $h= H J/D^2$. The functional \eqref{EnergyDensity} produces a rich phase diagram of magnetic phases, including a cone phase, a helicoid phase, isolated skyrmions, and skyrmion lattices \cite{Bogdanov94,Butenko10,Wilson14}. In the presence of additional interactions, such as Rashba spin-orbit coupling, the stability of skyrmion phases is enhanced over a large range of magnetic fields\cite{Banerjee14}. In the following we will consider the regime of isolated skyrmions which exists for a wide range of parameters $\kappa$ and $h$ as a metastable state of the ferromagnetic background $\mb{m}=(0,0,1)$. Rotationally symmetric solutions are described by
\begin{equation}
\Theta_0(\rho,\phi)=\Theta_0(\rho) , \qquad \Phi_0(\rho,\phi)= \phi + \pi/2 \,,
\label{SolTheta}
\end{equation}
for a skyrmion with topological number $Q=-1$. The skyrmion energy with respect to the uniform state is, 

\begin{eqnarray}
w_0(\Phi_0,\Pi_0)&=& (\nabla \Theta_0)^2 +(\frac{1}{\rho^2}+\kappa) \sin ^2\Theta_0 + h(1-\cos \Theta_0) \nonumber \\
 &+&\Theta_0' +\frac{1}{\rho}\sin \Theta_0 \cos \Theta_0 \,.
\end{eqnarray}
The structure of the stationary skyrmion is determined by the Euler--Lagrange equation,
\begin{eqnarray}
\Theta_0''(\rho)+\frac{\Theta_0'(\rho)}{\rho}&-&\frac{\sin\Theta_0 \cos\Theta_0 }{\rho^2} + \frac{\sin^2\Theta_0}{\rho} \nonumber \\
&-&\frac{h}{2} \sin\Theta_0 -\frac{\kappa}{2} \sin2\Theta_0=0\,,
\label{EulerEq}
\end{eqnarray}
with boundary conditions $\Theta_0(0)=\pi$ and $\Theta_0(\infty)=0$. Because there is no known analytic solution, the following function can be used as an approximate solution
\begin{equation}
\Theta_0(\rho)=2 \tan^{-1} \left( \frac{\lambda}{\rho}e^{-\frac{\rho -\lambda}{\rho_0}} \right)\,,
\label{SmallRadiusSolut}
\end{equation}
where $\rho_0=\sqrt{2/(2\kappa +h)}$. The skyrmion radius $\lambda$ can be determined by fitting the Ansatz \eqref{SmallRadiusSolut} to the one obtained numerically using an algorithm based on Runge--Kutta formulas \cite{Algorithm}. In the opposite limit of a large-radius skyrmion the magnetization profile is described by
\begin{equation}
\cos \Theta_0(\rho) = \tanh (\frac{\rho - \lambda}{\Delta_0}) \,,
\label{Profile_Lrg}
\end{equation} 
where the parameters $\lambda$ and $\Delta_0$ are again calculated numerically by fitting the approximate function \eqref{Profile_Lrg} to the numerical solution of the Euler-Lagrange equation \eqref{EulerEq}. Magnetization profiles of skyrmions in the small-radius limit are depicted in Fig.~\ref{Profile}.

\section{Magnon Spectrum}
\label{sec:ap2}

In this section we provide a full analysis of the calculation of magnon modes, including some general properties of the magnon Hamiltonian. 
\subsection{Non-Hermiticity}
Magnon modes are solution of the eigenvalue problem $\mc{H} \Psi^{R}_{n,a} = 2 \varepsilon_{n} \sigma_z \Psi^{R}_{n,a}$, where the explicit form of the Hermitian Hamiltonian $\mc{H}$ is given in Eq.~\eqref{EVPTransfm} below. Here, the index $R$ denotes the fact that $\Psi^{R}_{n,a}$ are right eigenvectors of the non-Hermitian matrix $\sigma_z \mc{H}$. Further, $n$ is the quantum number and $a$ counts for any possible degeneracy. Note that the pseudo-Hermitian operator $\sigma_z \mc{H}$ satisfies the relation 
\begin{align}
(\sigma_z \mc{H})^{\dagger} = \sigma_z (\sigma_z \mc{H}) \sigma_z \,.
\label{pseudo-Herm}
\end{align}
In addition to the right eigenstates $\Psi^{R}_{n,a}$ of operator $\sigma_z \mc{H}$, we introduce left eigenstates of the Hermitian adjoint matrix $(\sigma_z \mc{H})^{\dagger} \Psi^{L}_{n,a} = 2 \varepsilon_n^{*}  \Psi^{L}_{n,a}$. Hermiticity of the operator $\sigma_z \mc{H}$ is assured with respect to a new inner product between left and right eigenstates $\langle \Psi^{L}_{n,a} \vert \Psi^{R}_{m,a} \rangle = \delta_{n,m}$\cite{Gardas16}. Using Eq.\eqref{pseudo-Herm}, it can be shown that left and right eigenvectors are connected through the relationship $\langle \Psi^{L}_{n,a} \vert = \langle \Psi^{R}_{n,a}\vert  \sigma_z$. Therefore, biorthogonality conditions are enforced as
\begin{align}
\langle \Psi_{n,a} \vert \sigma_z \vert \Psi_{m,a} \rangle =\delta_{n,m} \,,
\end{align}
where we have now dropped the index $R/L$ which distinguish between right and left eigenfunctions. The identity operator is given by 
\begin{equation}
\mathbb{1} = \sum_{n,a} \vert \Psi_{n,a} \rangle \langle \Psi_{n,a} \vert \sigma_z \,,
\label{UnityOperator}
\end{equation}
while the trace of an operator $A$ is computed as
\begin{align}
\mbox{Tr}(A) = \sum_{n,a} \langle \Psi_{n,a} \vert \sigma_z  A  \vert \Psi_{n,a} \rangle \,.
\end{align}
\subsection{Particle-Antiparticle symmetry}
\label{sec:ap2ii}

In this section we discuss a particular symmetry for the magnon Hamiltonian $\mc{H}$ which is known for the case of magnetic excitations in ferromagnetic 2D skyrmions \cite{SchuttePRB90} and for vortices in 2D easy-plane ferromagnets \cite{Sheka04}. In particular, there exist a transformation matrix $C=K\sigma_x$, where $K$ denotes complex conjugation, under which Hamiltonian $\mc{H}$ is invariant $C \mc{H} C^{\dagger} = \mc{H}$, while the Pauli matrix $\sigma_z$ changes sign $C \sigma_z C^{\dagger} =-\sigma_z$. This symmetry generates an additional class of solutions of the same eigenvalue problem, but with negative eigenfrequencies. 

More specifically, it can be shown that if states $\Psi_{n,a}$ are solutions of the eigenvalue problem (EVP)
\begin{align}
\mc{H} \Psi_{n,a} = 2 \varepsilon_n \sigma_z \Psi_{n,a}
\label{EVP}
\end{align} 
with eigenfrequencies $\varepsilon_n \geq 0$, then states $ C \Psi_{n,a} $ are solutions of the same EVP \eqref{EVP} with eigenfrequency $-\varepsilon_{n}$. Such a symmetry is called \textit{particle-antiparticle}. To describe these two classes of solutions, we introduce the index $s=\pm 1$ such that $\Psi^{1}_n$ with eigenfrequency $\varepsilon_{n}^{1}=+\varepsilon_n $ corresponds to particles, while $\Psi^{-1}_{n,a} =C \Psi^{1}_{n,a}$ with $\varepsilon_n^{-1}=-\varepsilon_n $ to antiparticles. The biorthogonality conditions for the solutions $\Psi^{s}_{n,a}$ are
\begin{align}
\langle \Psi^{s}_{n,a} \vert\sigma_z \vert \Psi^{s}_{m,a} \rangle &= s \delta_{n,m} \nonumber \\
\langle \Psi^{s}_{n,a} \vert \sigma_z \vert \Psi^{-s}_{m,a} \rangle &= 0 \,. \label{Orth}
\end{align}
From the conditions \eqref{Orth}, it follows that the energy levels are not necessarily linked to the eigenfrequencies $\varepsilon_n$. The measure of the energy of a mode $\Psi_{n,a}$ is given by the quantity
\begin{equation}
E_n =\langle \Psi^{s}_{n,a} \vert \mc{H} \vert \Psi^{s}_{n,a} \rangle = \varepsilon^{s}_n \langle \Psi^{s}_{n,a}\vert  \sigma_z \vert \Psi^{s}_{n,a} \rangle  \,.
\label{Energylevels}
\end{equation}
It is easily demonstrated that modes with eigenenergies $\varepsilon_n$ and $-\varepsilon_n$ have their energies $E_n$ equal in sign (positive) and value. The unity operator is redefined as
\begin{align}
\mathbb{1} = \sum_{s=\pm 1} \sum_{n,a} s \vert \Psi^{s}_{n,a} \rangle \langle \Psi^{s}_{n,a} \vert \sigma_z  \,,
\end{align}
and the trace of an operator changes accordingly,
\begin{align}
\mbox{Tr}(A) = \sum_{s=\pm 1} \sum_{n,a} s  \langle \Psi^{s}_{n,a} \vert \sigma_z  A \vert \Psi^{s}_{n,a} \rangle \,.
\end{align}

\subsection{Eigenvalue problem}
Here we consider the spectrum of the magnon modes on the skyrmion background. For reasons of convenience, in this section we use the notation $\Psi_n$ to denote particle states, and antiparticles states are recovered using the particle-antiparticle symmetry described in the preceding subsection. Magnon scattering states are obtained for energies $\varepsilon_n \geq \varepsilon_{\mbox{\tiny{MS}}}$, with $\varepsilon_{\mbox{\tiny{MS}}}=\kappa+h/2$. In addition to scattering states, we expect localized modes that correspond to deformations of the skyrmion into polygons (breathing modes) in the range $0 < \varepsilon_n < \varepsilon_{\mbox{\tiny{MS}}}$. The existence of such modes was found numerically in Refs. [\onlinecite{LinPRB14}] and [\onlinecite{SchuttePRB14}]. An extensive analysis of the magnon spectrum has been provided in Ref.~[\onlinecite{SchuttePRB14}], where the authors numerically determine the magnon bound states and  provide an analytical and numerical description of magnon scattering states. Here we repeat some steps of the analysis adapted to our puprose and introduce a variational approach for the calculation of the bound states in the small and large radius limit of skyrmions.

To begin with, we seek for solutions of the EVP \eqref{EVP} , where the Hamiltonian  is given by
\begin{equation}
\mc{H}=2[- \nabla^2 + U_0(\rho)] \mathds{1} + 2 W(\rho) \sx - 2 i V(\rho) \frac{\pt}{\pt \phi}\sz \,.
\label{EVPTransf}
\end{equation}
Here, the potential terms $V(\rho)$, $W(\rho)$, and $U_0(\rho)$ are defined by
\begin{align}
V(\rho)&=\frac{2\cos \Theta_0}{\rho^2} -\frac{ \sin \Theta_0}{\rho} \,, \nonumber \\
W(\rho)&=\frac{\sin 2\Theta_0 }{4\rho}- \frac{(\Theta_0')^2}{2}+ \frac{1}{2}(\kappa+ \frac{1}{\rho^2}) \sin^2 \Theta_0 - \frac{ \Theta_0'}{2} \,,
\end{align}
and 
\begin{align}
U_0(\rho)= \frac{h \cos \Theta_0}{2}  -\frac{3\sin 2 \Theta_0 }{4 \rho}   -\frac{(\Theta_0')^2}{2} \nonumber \\
+(\frac{\kappa}{4}+\frac{1}{4 \rho^2}) (1+ 3 \cos 2\Theta_0)-\frac{\Theta_0'}{2} \,, 
\end{align}
where the quantum number $n$ labels both localized modes and scattering states. Next, we represent solutions in terms of wave expansions $\Psi_{n}=e^{i  m \phi}\psi_{n,m}(\rho)$, and the EVP is written as $\mc{H}_m  \psi_{n,m}(\rho) = \varepsilon_{n,m} \sz  \psi_{n,m}(\rho)$ with 
\begin{equation}
\mc{H}_m =  (- \nabla^2_{\rho}+ U_0(\rho) +\frac{ m^2}{\rho^2} ) \mathds{1}  
+ V(\rho) m \sz + W(\rho) \sx\,,
\label{EVPTransfm}
\end{equation}
where $\nabla^2_{\rho}= \frac{\pt^2}{\pt_{\rho}}+\frac{1}{\rho} \frac{\pt}{\pt_{\rho}}$. These eigenfunctions are normalized such that
\begin{equation}
\int_0^{\infty} d \rho~ \rho \psi_{n,m}^{\dagger} (\rho)\sz \psi_{n',m}(\rho) =\delta_{n,n'} \,.
\label{Normaliz1}
\end{equation}
One of the translational modes with zero energy is given by 
\begin{equation}
\Psi^{1}_1= e^{i\phi} \frac{1}{\sqrt{8}}\binom{\Theta_0' -\frac{1}{\rho} \sin \Theta_0}{\Theta_0' +\frac{1}{\rho} \sin \Theta_0}\,
\label{ZMode1}
\end{equation}
and the other is found using the particle-antiparticle symmetry
\begin{equation}
\Psi^{-1}_{1}= K \sx \Psi^{1}_1 = e^{-i\phi} \frac{1}{\sqrt{8}}\binom{\Theta_0' +\frac{1}{\rho} \sin \Theta_0}{\Theta_0' -\frac{1}{\rho} \sin \Theta_0} \,.
\label{ZMode2}
\end{equation}
Here, the upper index corresponds to particle/antiparticle while the lower to the quantum number $m$. 
\begin{figure}[b]
\centering
\includegraphics[width=1\linewidth]{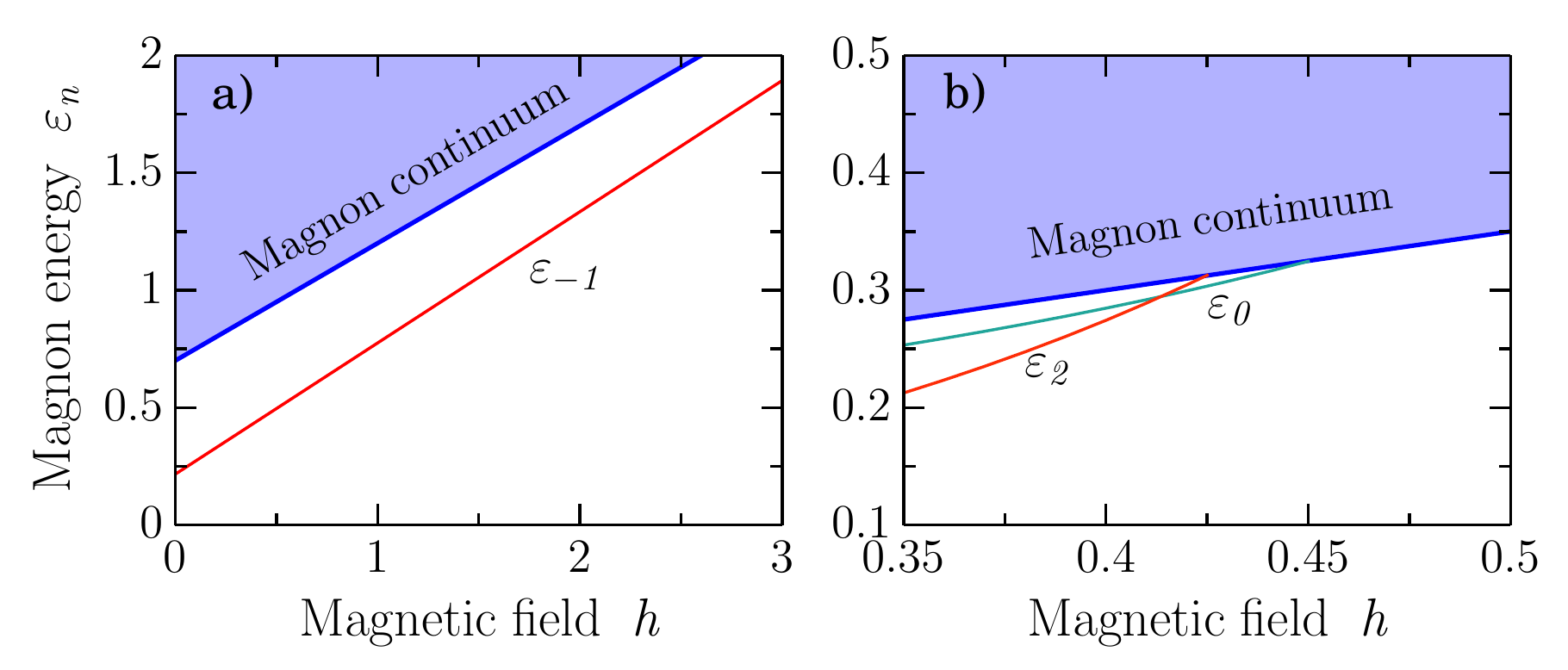}
\caption{\ Magnetic field dependence of the energy of the localized mode with (a) $m=-1$ for a skyrmion with $\kappa=0.7$ described by Eq.~\eqref{Psim1} and (b) with $m=0$ and $m=2$ for a skyrmion with $\kappa=0.1$ described by Eq.~\eqref{BS_LL}. Shaded areas depict the magnon continuum with boundary line $\varepsilon_{\mbox{\tiny{MS}}}$.
} \label{MagnonEn}
\end{figure}

\subsubsection{Small-Radius Skyrmion}
\label{subsec:SmallR}

For magnetic solitons in $2D$ easy--axis ferromagnets in the small--radius limit there is a bound state with $m=-1$ that remains localized and is associated with the soliton displacement \cite{Ivanov05}. We extend these results to the skyrmion field, and we seek for a localized mode with $m=-1$ by performing a variational calculation. We choose trial functions $\Psi_{-1}=e^{-i \phi}\psi_{-1}(\rho)$ as \cite{Ivanov05}
\begin{equation}
\psi_{-1}(\rho)= \binom{\Theta_0^{'} -\sin \Theta_0/\rho + a \rho^2 \Theta_0 ^{'} - b \rho \sin \Theta_0}{\Theta_0^{'} -\sin \Theta_0/\rho + a \rho^2 \Theta_0 ^{'} +b \rho \sin \Theta_0} \,,
\label{Psim1}
\end{equation}
and minimize the functional
\begin{equation}
\mathcal{U}= \int d\rho ~\rho~ \psi_{-1}^{\dagger}(\rho) (\mc{H}_{-1}- \varepsilon_{-1} \sz) \psi_{-1}(\rho)  \,
\end{equation}
with respect to the variational parameters $a$ and $b$ in order to calculate the energy $\varepsilon_{-1}$. Then, the parameters $a$ and $b$ can be found from the normalization conditions \eqref{Normaliz1}. In Fig.~\ref{MagnonEn}~(a) we illustrate the energy $\varepsilon_{-1}$ as a function of magnetic field $h$ for $\kappa=0.7$. Within the limits of the variational approach, we find $\varepsilon_{-1} < \varepsilon_{\mbox{\tiny{MS}}}$ for a wide range of magnetic fields up to $h=8$. 

\subsubsection{Large-Radius Skyrmion}
\label{subsec:largeradius}
In the limit of large radius, the magnetization profile of the skyrmion is described by Eq.~\eqref{Profile_Lrg} and the localized modes $\Pbs_m$ with energy $\varepsilon_m$ are again found variationally, but now using the trial functions \cite{Sheka01}
\begin{align}
\Pbs_m(\mb{r}) = \frac{1}{\sqrt{2 \pi}} e^{i m \phi} \binom{a_m f_0(\rho)}{b_m f_0(\rho)}\,,
\label{BS_LL}
\end{align}
where $f_0(\rho)=A/\cosh(\frac{\rho-\lambda}{\Delta_0})$ and $A$ is chosen such that $\int d\rho \rho f_0^2 =1$. The condition of minimizing the energy functional $\mc{U}=\int d\mb{r} (\Pbs_m)^{\dagger} (\mc{H}_m - \sz \varepsilon_m) \Pbs_m $, along with normalization conditions satisfied by the functions $\Pbs_m(\rho)$ will specify the eigenenergies $\varepsilon_m$ as well as the variational parameters $a_m$ and $b_m$. In Fig.~\ref{MagnonEn}~(b) we display the energies $\varepsilon_{0}$ and $\varepsilon_{2}$ as a function of magnetic field $h$ for $\kappa=0.1$.

\subsubsection{Scattering States}
\label{subsec:ScattStates}
To complete the description of the magnon spectrum we need to include the scattering states $\Psc_{m,k}(\mb{r})$ classified by $m$ as well as the radial momentum $k \geqslant 0$. Here we repeat some steps of the calculation of $\Psc_{m,k}(\mb{r})$ originally presented in Ref.~[\onlinecite{SchuttePRB14}] for reasons of a complete discussion. In the absence of a skyrmion, magnon scattering states are described by
\begin{align}
\mc{H}^0_m =  (- \nabla^2_{\rho}+ \frac{ m^2+1}{\rho^2} +\ems ) \mathds{1}  +\frac{2 m}{\rho^2} \sz \,,
\end{align}
and the eigenvalue problem $\mc{H}^0_m \pfr_{m,k} = \varepsilon_k \pfr_{m,k}$  is solved by
\begin{align}
\pfr_{m,k} =\nu_m J_{m+1}(k \rho) \binom{1}{0} \,,
\label{SCfree}
\end{align} 
where $J_{m}$ are the Bessel functions of the first type, $\nu_m$ a normalization constant and $\varepsilon_k= \ems+k^2$, with $\ems = \kappa+h/2$. In the presence of a skyrmion the scattering problem of Eq.~\eqref{EVPTransfm} is reformulated as $\mc{H}_m=\mc{H}^0_m + \Vsc$ with
\begin{align}
\Vsc = u_0(\rho)  \mathds{1} +\left[ V(\rho) - \frac{2}{\rho^2} \right] m \sz + W(\rho) \sx\,,
\end{align} 
with potential $u_0(\rho)= U_0(\rho) -\frac{h}{2} -\frac{1}{\rho^2} - \kappa $. In this case, the solutions of the eigenvalue problem $\mc{H}_m \psc_{m,k} = \varepsilon_k \psc_{m,k}$ are 
\begin{equation}
\psc_{m,k}(\rho)= d_m \left[\cos(\delta_m) J_{m+1}(k \rho)- \sin(\delta_m) Y_{m+1}(k \rho) \right] \binom{1}{0}\,,
\label{SC}
\end{equation}
where $Y_m$ are the Bessel functions of the second kind, $d_m(k)$ is a normalization constant and $\delta_m(k)$ is a scattering phase shift that determines the intensity of magnon scattering due to the presence of the skyrmion. The phase shifts can be calculated within the WKB approximation \cite{SchuttePRB14, Berry72} as
\begin{align}
\delta_m(k)=& \int_{\rho_1}^{\infty} d\rho \left(\sqrt{k^2 + \ems - \Uwkb(\rho)} -k \right) d\rho \nonumber \\ &+\frac{\pi}{2} \vert m+1 \vert - k \rho_1 \,, 
\label{PhaseSh}
\end{align}
where $\rho_1$ is the first classical turning point $\varepsilon_k = \Uwkb(\rho_1)$, when approaching $\Uwkb(\rho)$ from large distances. The effective WKB potential is given by
\begin{align}
\Uwkb(\rho)= \varepsilon_k - \frac{\mu( \rho/\rho_0)}{\rho^2} \,,
\end{align}
with $\rho_0= \sqrt{2/(2 \kappa+ h)} =1/\sqrt{\ems}$ and $\mu(x)$ is the eigenvalue of the operator
\begin{align}
\Lambda(x) =-(m\mathds{1}+\sz)^2 + \rho_0^2 e^{2 x} (-\frac{1}{\rho_0^2}\mathds{1}+\varepsilon_k \sz -\Vsc (\rho_0 e^{x})) \,,
\end{align}
which corresponds to the eigenvector $v(x) \propto \binom{1}{0}$. Here, we use $x= \log(\rho/\rho_0)$ and $v(x)=\psc_m(\rho_0 e^x)$. Taking into account the angular coordinate, scattering states are described by
\begin{align}
\Psc_{m,k}(\mb{r}) =\frac{e^{i m \phi}}{\sqrt{2 \pi}}\psc_{m,k}(\rho) \,,
\label{SCm}
\end{align}
where $\psc_{m,k}(\rho)$ is given in Eq.~\eqref{SC}, while scattering phase shifts $\delta_m(k)$ are calculated with the help of Eq.~\eqref{PhaseSh}. 
\section{Technical Details on the Mass Tensor}
\label{sec:ap3}

In this Appendix we discuss various technical details relevant to the mass tensor $\mc{M}^0_{ij}$ and $\mc{M}^T_{ij}$. 

\emph{First order perturbation theory for the calculation of mass term $\mc{M}^0_{ij}$}. To investigate the effect of the external perturbations to magnon modes it is convenient to consider small perturbations $\mc{V}$ that alter the eigenfunctions $\tilde{\Psi}_n = \Psi_n + \delta \Psi_n$, where $ \Psi_n$ are eigenfunctions of the unperturbed Hamiltonian $\mc{H}$ given in Eq.~\eqref{EVPTransf}. Since damping and mass depend on the inverse of the magnon energy $\varepsilon_n$, we expect that bound states below the gap have a larger contribution to the mass terms $\mc{M}^0_{xx}$ and $\mc{M}^0_{yy}$ appearing in Eq.~\eqref{MassTerm}. Among them, the state $\Psi^{s}_{-1}$ turns out to be the most important one for the form of external perturbations considered here. Under the assumption that we keep only the state $\Psi^{s}_{-1}$ in the sum over all magnon modes we arrive at 
\begin{align}
\mc{M}^{0}_{ii}=   N_A \bar{S} \sum_{s,s'=\pm1} \vert D_{s,s'} \vert^2 +N_A \bar{S} l_{i} \sum_{s=\pm 1}(D_{s,s}D^{*}_{-s,s}+c.c.) 
\end{align}
where $l_x=1$, $l_y=-1$ and $D_{s,s'} = \langle \Psi_{1}^{s} \vert \mc{W} \vert \Psi_{-1}^{s'} \rangle$.  
Off-diagonal elements $\mc{M}_{xy}$ and $\mc{M}_{yx}$ do not contribute to the equation of motion for the skyrmion center-of-mass $\mathbf{R}$. In the above expressions we have used the following exact relations,
\begin{align}
f_x= -\frac{i}{\sqrt{2}} \sz (\Psi^{1}_{1}+\Psi^{-1}_{1}) \,, \\
 f_y= -\frac{1}{\sqrt{2}} \sz (\Psi^{1}_{1}-\Psi^{-1}_{1}) \,.
\end{align}
The eigenstate $\Psi^{s}_{-1}$ has been calculated variationally in Appendix \ref{sec:ap2} in the small--skyrmion limit. For this calculation, the skyrmionic static configuration is required, where we use solution \eqref{Solution} and \eqref{SmallRadiusSolut} for the angle $\Theta_0(\rho)$ for a given set of parameters $\kappa$ and $h$. The skyrmion radius $\lambda$ is determined by fitting the trial function \eqref{SmallRadiusSolut} to the one obtained numerically. For all external perturbations we have considered in Sec. \ref{sec:Chiral Magnets}, we find that $\mc{M}^0_{xx}=\mc{M}^0_{yy}$, whenever $\mc{V}$ is isotropic in $\hat{x}$ and $\hat{y}$ direction. 

\emph{Finite temperature mass term $\mc{M}^T_{ij}$}. From the angular dependence of the eigenstates $\Psi_n(\phi, \rho) = e^{i m \phi} \psi_{n,m} (\rho)$ and the form of the operator $\Gamma_i$ given in Eq.~\eqref{OperatorKJ} we conclude that in the finite $T$ expression of the mass Eq.~\eqref{MassT} only states with angular momentum difference $\Delta m = \pm 1$ contribute. 
For the parameter range considered here we find two bound states below the continuum with $m=0$ and $m=2$. Since the zero mode $m=1$ is excluded from summations, there are no bound--bound state contributions. To calculate the contribution from the scattering states, the sum over the quantum number $n$ in Eq.~\eqref{MassT} is replaced by 
\begin{align}
\sum_{n} \Psi_n \rightarrow \sum_{m} \sum_{k} \Psc_{m,k} \rightarrow \sum_{m} \left( \sum_k \Pfr_{m,k}  - \sum_{k}  \Psc_{m,k}\right), 
\end{align}
where we have subtracted background fluctuations described by Eq.~\eqref{SCfree} in order to render the result finite in the thermodynamic limit. In addition, the discrete sum over $k$ is replaced by $\sum_k \rightarrow \frac{L}{\pi} \int_0^{\infty} dk ~k$. 

Next, we consider the particle-particle expression for $\Mbscx$,
\begin{align}
\Mbsci= \sum_{m} \int_0^{\infty} dk~ k \frac{\mc{C}_{m,k} }{\bar{\varepsilon}_m - \bar{\varepsilon}_k} [\coth(\frac{\beta \bar{\varepsilon}_k}{2}) - \coth(\frac{\beta \bar{\varepsilon}_m}{2})] \,,
\end{align}  
with elements $\mc{C}_{m,k}= \frac{1}{8} \tilde{c}_m^2 (g_{m,m-1}+g_{m,m+1})$, and 
 \begin{align}
 g_1^{m,m'}=(1- \cos(\delta_{m'})^2) &\left[ \frac{a_m k}{2}(D_1^{m'} -D_1^{m'+2} ) \right. \nonumber \\
&  \left.  +a_m m' D_2^{m'+1}+ b_m D_3^{m'+1} \right]^2 \,,
 \end{align}
with integrals $D_1^{m}= ( f_0 J_{m} )$, $D_2^{m}= ( f_0 J_{m} /\rho )$ and $D_3^{m}= ( f_0 J_{m} \Theta_0' \cot \Theta_0 )$. We use $( \cdots ) = \int_0^{\infty}  \cdots d\rho \rho$. The normalization constant $\tilde{c}_m$ is defined as 
\begin{equation}
\tilde{c}_m (k) = \frac{\sqrt{L/\pi}}{\sqrt{\int_0^{L} d\rho \rho J^2_{m+1}(k \rho)}}\,,	
\end{equation}
for a finite system size of area $\pi L^2$. Note that we have neglected terms proportional to $\sin(\delta_m)$ under the assumption $\cos(\delta_m) \gg \sin(\delta_m)$. In Fig.~\ref{QMass}-(a) we present the magnetic field dependence of $\Mbsci$ for $T=0.05$ for a skyrmion with $\kappa=0.1$, $S=1$, $\tilde{J}/\tilde{D}=4$ and $L \rightarrow \infty$. For this choice, we take into account two localized modes with energy less than $\ems$ in the magnetic field region between $h=0.34$, where the skyrmion energy becomes positive, up to $h=0.42$ where the bound state energy passes into to the continuous spectrum (see Fig.~\ref{MagnonEn}~(b)).  
To calculate the particle-particle contribution from the scattering states, we use the states $\Psc_{m,k}$ given in Eq.~\eqref{SC}, and under the assumption that $\cos^2 (\delta_m(k)) \gg  \sin^2 (\delta_m(k)) $ we arrive at the following result,
\begin{align}
&\Msci=\frac{1}{8} \sum_{m} \int_0^{\infty} dk \frac{\beta k^2 g_{m,m+1}^{k,k} }{\sinh^2(\beta \bar{\varepsilon}_k/2)} \nonumber \\
+&\frac{1}{8} \sum_{m>0} \int_0^{\infty} k dk \int_{k}^{\infty} k^{'} d k^{'} \frac{C_{k,k'}(\beta)}{\bar{\varepsilon}_k - \bar{\varepsilon}_{k'}} ( G_{m,m+1}^{k,k'}+ G_{m,m-1}^{k,k'} )\,,
\end{align}
with $g_{m,m'}^{k,k'} = \tilde{c}_m (k)^2 \tilde{c}_m (k')^2  [-\cos^2 (\delta_m(k))  \cos^2 (\delta_{m'}(k'))+1]$, $C_{k,k'}(\beta)= [\coth(\beta \bar{\varepsilon}_{k'}/2) -\coth(\beta \bar{\varepsilon}_{k}/2)] $ and $G_{m,m'}^{k,k'}= g_{m,m'}^{k,k'} k^{(m +m'+1)} (k')^{-(m+m'+3)} $. In Fig.~\ref{QMass}-(b) we plot the temperature and magnetic field dependence of $\Msci$ for a skyrmion with $\kappa=0.1$, $S=1$, $\tilde{J}/\tilde{D}=4$ and $L \rightarrow \infty$. Summations over quantum number $m$ converge and are bounded between $-7 \leqslant m  \leqslant7$. The particle-particle and antiparticle-antiparticle contribution to the effective mass are equal, since operator $\Gamma_i$ is invariant under the particle-antiparticle symmetry. Finally, the particle-antiparticle contribution to the effective mass is calculated in a similar way.


\begin{thebibliography}{99}
\bibitem{NoteN_A} The quasi 2D regime with a homogeneous distribution along the $z$ axis is valid as long as the transverse to the magnetic layer spin waves around the skyrmion are frozen. These modes acquire a finite size gap for a finite sample $t$ and are frozen out below a given temperature for sample widths $t < t_{cr}$, where $t_{cr}=\pi \frac{\tilde{J}}{\tilde{D}} \sqrt{\frac{J}{k_B T}}$. Note that below a second critical thickness $t'_{cr}$, a skyrmion gains additional 3D conical modulations along z axis \cite{Rybakov13}, a twist which mostly affects the magnetic spins near the surfaces. Typical numbers for the magnetic insulator Cu$_2$OSeO$_3$ are $t_{cr}/\alpha= 90$ layers at $T$=1 K and $t'_{cr}/\alpha\simeq 15$ layers.
\end{thebibliography}
\end{document}